\documentclass[times,twocolumn,final]{elsarticle}

\makeatletter
\def\ps@pprintTitle{%
 \let\@oddhead\@empty
 \let\@evenhead\@empty
 \def\@oddfoot{}%
 \let\@evenfoot\@oddfoot}
\makeatother

\usepackage{framed,multirow}
\usepackage{amssymb}
\usepackage{amsmath}
\usepackage{latexsym}
\usepackage{hyperref}
\usepackage{siunitx}
\usepackage[capitalise]{cleveref}
\usepackage{booktabs}
\usepackage{siunitx}

\begin{document}

\begin{frontmatter}

\title{Atlas-ISTN: Joint Segmentation, Registration and Atlas Construction with Image-and-Spatial Transformer Networks}

\author[a,b]{Matthew Sinclair\corref{cor1}}
\cortext[cor1]{Corresponding author:}
\ead{msinclair@heartflow.com}
\author[a,b]{Andreas Schuh}
\author[a,b]{Karl Hahn}
\author[a]{Kersten Petersen}
\author[a]{\\Ying Bai}
\author[a,b]{James Batten}
\author[a,b]{Michiel Schaap}
\author[a,b]{Ben Glocker}

\address[a]{HeartFlow, USA}
\address[b]{Biomedical Image Analysis Group, Imperial College London, UK}

\begin{abstract}
Deep learning models for semantic segmentation are able to learn powerful representations for pixel-wise predictions, but are sensitive to noise at test time and do not guarantee a plausible topology. Image registration models on the other hand are able to warp known topologies to target images as a means of segmentation, but typically require large amounts of training data, and have not widely been benchmarked against pixel-wise segmentation models. We propose Atlas-ISTN, a framework that jointly learns segmentation and registration on 2D and 3D image data, and constructs a population-derived atlas in the process. Atlas-ISTN learns to segment multiple structures of interest and to register the constructed, topologically consistent atlas labelmap to an intermediate pixel-wise segmentation. Additionally, Atlas-ISTN allows for test time refinement of the model's parameters to optimize the alignment of the atlas labelmap to an intermediate pixel-wise segmentation. This process both mitigates for noise in the target image that can result in spurious pixel-wise predictions, as well as improves upon the one-pass prediction of the model. Benefits of the Atlas-ISTN framework are demonstrated qualitatively and quantitatively on 2D synthetic data and 3D cardiac computed tomography and brain magnetic resonance image data, out-performing both segmentation and registration baseline models. Atlas-ISTN also provides inter-subject correspondence of the structures of interest, enabling population-level shape and motion analysis. 
\end{abstract}

\end{frontmatter}

\section{Introduction}

\begin{figure*}[!ht]
    \centering
    \includegraphics[width=0.9\linewidth]{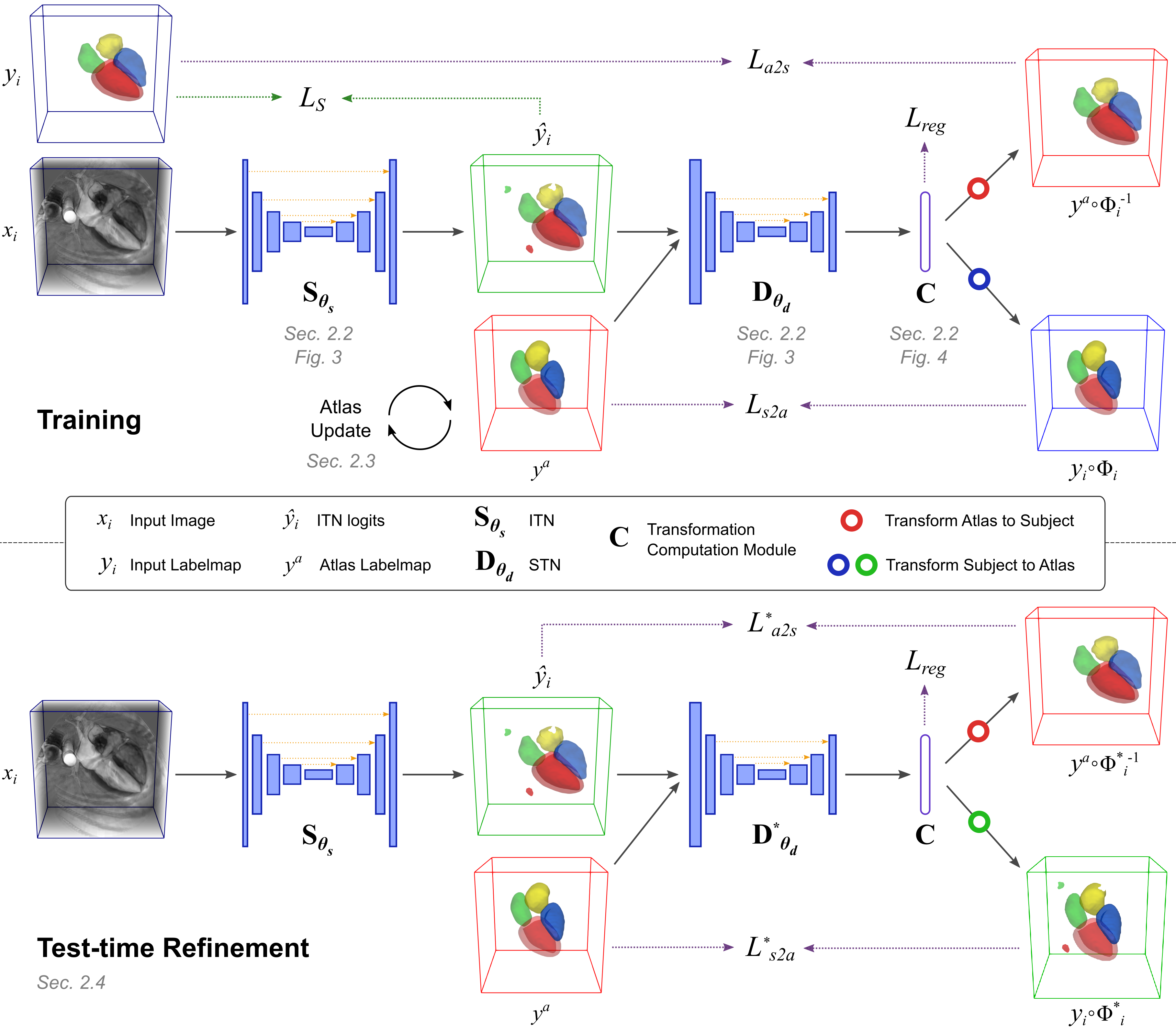}
    \caption{An overview of the Atlas-ISTN framework in training (top) and test time refinement (bottom) illustrated with CCTA image data. During training, both ITN and STN weights are optimized by leveraging ground-truth labelmaps for each subject, while the atlas is updated at the end of each epoch. A symmetric loss is used to register the atlas to the subject and vice versa. At test time, instance-specific optimization (indicated by *) leverages the ITN prediction as a target labelmap to update the STN weights, providing a refined transformation from the atlas to a given subject. Spurious segmentations in the ITN prediction can be circumvented via registration of the atlas to the subject. References to related sections and figures are provided in grey text. Red boxes: atlas labelmap (deformed and undeformed). Blue boxes: input image or corresponding GT labelmap (deformed and undeformed). Green boxes: ITN logits (deformed and undeformed).}
    \label{fig:framework}
\end{figure*}

\paragraph{Motivation}
Image segmentation and registration have long been important tools for biomedical image analysis \citep{Maintz1998,Pham2000}. Deep learning models such as U-nets \citep{Ronneberger2015} have emerged as the state-of-the-art for segmentation, with their ability to learn rich feature representations for accurate pixel-wise segmentations on challenging image datasets when trained with large labelled sets of 2D and 3D images. One challenge with such segmentation models however is their sensitivity to image noise and artefacts, which can yield spurious and topologically implausible segmentations at test time. Furthermore, such undesirable predictions are made more likely with fewer training examples. Numerous recent works have sought to tackle this with post-processing \citep{Kamnitsas2017,Larrazabal2019}, anatomical constraints in training \citep{Oktay2018}, and novel regularizers or loss terms \citep{Xu2019,Clough2019}, to name a few. While such approaches have improved topological plausibility of segmentation model predictions, none guarantee a target topology at test time.  

More recently, a growing body of research has explored the use of deep learning models for image registration for the purpose of image segmentation. Deep learning registration models typically learn to predict a dense deformation field to register a pair of images, which can be used to propagate a labelmap (of known topology) from a source to a target image for the purpose of segmentation. Most such methods rely on a single pass of a trained model at test time to predict a deformation field \citep{Dalca2018a,Lee2019a,Balakrishnan2018,Dalca2019,Dalca2019b,Dong2020,Mansilla2020}. The accuracy of a warped segmentation to a target image has been shown to improve with test time optimization of the registration network's parameters, particularly in settings of limited training data \citep{Balakrishnan2018,Lee2019b}. \cite{Dalca2019} proposed a framework to learn a conditional atlas image jointly with a model to perform registration of the atlas with target images. These recent image-registration driven approaches used for segmentation are however not commonly benchmarked against pixel-wise segmentation models, and have fallen short in relevant metrics when they have been \citep{Lee2019a,Xu2019}.

Inspired by features of each of these approaches, we propose Atlas-ISTN, a framework that benefits from the detailed predictions of pixel-wise segmentation while circumventing the effects of noise and artifacts via registration of a learned topologically consistent, population-derived multi-class atlas labelmap (Fig. \ref{fig:framework}, upper panel) to an initial pixel-wise segmentation. Additionally, Atlas-ISTN leverages test time refinement of model parameters to optimize for the registration of the constructed atlas labelmap to a predicted segmentation labelmap, improving over 1-pass test time performance (Fig. \ref{fig:framework}, lower panel). This framework also simultaneously guarantees topology of the structures-of-interest (SoI) and provides atlas-space correspondence between subjects for further population-level analysis, all while being contained in a single model framework with straight-forward training and test time deployment.

\paragraph{Related work}

Medical image segmentation models can be broadly categorized into pixel-wise prediction, shape fitting and registration-based methods. A rich body of literature exists for methods which use image registration as a means to segment a target image, which build on two main approaches: multi-atlas segmentation (MAS) \citep{Isgum2009,Kirisli2010,Iglesias2015}, and construction and registration of a statistical shape model (SSM) \citep{Heimann2009, Young2009}. MAS has proven to be highly effective, and provides competitive performance with modern deep learning methods for segmenting large structures of the heart from 3D cardiac computed tomographic angiography (CCTA) and magnetic resonance imaging (MRI), albeit in a setting with limited training data \citep{Heinrich2018,Zhuang2019}. MAS has also been used effectively for segmentation of heart structures in CCTA in a large-scale multicenter/multivendor evaluation \citep{Kirisli2010}. A down-side of MAS is the high computational overhead at test time, which can involve registration, selection and sophisticated fusion of multiple atlas labelmaps to a target image to achieve best performance \citep{Isgum2009}. 

Methods using SSMs on the other hand typically construct a template/atlas in the form of a mean image, labelmap, or mesh, which at test time is registered to a target image, with the option of leveraging a segmentation of the target image in the registration process \citep{Medrano-Gracia2014,Bai2015}. Advantages of SSMs include providing correspondence to a common atlas space for population-level analyses of shape and motion, and using an unbiased atlas tends to perform better for segmentation than warping a given training case. For cardiac data, given the significant variability in heart orientation, size and morphology observed in CCTA and cardiac MRI, a two-stage (i) affine and (ii) non-rigid registration approach is typically used, sometimes requiring the definition of anatomical landmarks to guide the first stage of registration \citep{Bai2015}. SSMs are also often parameterized with Principal Component Analysis (PCA), and the PCA modes can be optimized to fit a SSM to a target image or segmentation \citep{Heimann2009}. Limitations of PCA representations however include over-fitting of the model to limited training data, thus not being able to accurately represent anatomies which lie outside of the training distribution and can include both significant (large-scale) and subtle (small-scale) variations in target anatomies. Both MAS and SSM-based segmentation approaches often involve workflows with multiple processing steps. For example, separate tools are typically used to optionally (1) build a SSM, (2) produce an image segmentation or landmark coordinates from an unseen target image, (3) register the atlas to a target image, segmentation, and/or landmarks (4) select and/or fuse the best atlas labelmap(s). See \citep{Iglesias2015} for a review of MAS and \citep{Heimann2009,Young2009} for a review of SSM works. 

Prior to the emergence of powerful convolutional neural networks (CNNs) for pixel-wise segmentation \citep{Ronneberger2015, Long2015}, MAS and SSM-based methods were the dominant approaches used for biomedical image segmentation of structures known to adhere to a particular atlas geometry. An advantage of the more traditional approaches is the preservation of topology, where pixel-wise deep learning segmentation models such as the U-net \citep{Ronneberger2015} or fully convolutional network \citep{Long2015} can suffer from spurious and anatomically implausible segmentations at test time. Deep learning segmentation models typically improve with more training data, but are still prone to errors due to domain shift and out-of-distribution test cases (e.g. originating from different scanners, sites, acquisition protocols and caused by imaging artifacts). Among the many methods that have been proposed to tackle these challenges, we summarize a few that focus on reducing spurious segmentations and encourage anatomically plausible predictions. Firstly, post-processing steps have been commonly used, such as fully-connected conditional random fields (CRF) and connected component analysis \citep{Kamnitsas2017}, as well as shape-aware denoising auto-encoders \citep{Larrazabal2019}. Other approaches include anatomically constrained neural networks (ACNNs) where shape-regularization is enforced with a latent representation of the underlying anatomy via an autoencoder \citep{Oktay2018,Chen2019}. One approach leveraged shape priors during training by using smooth 3D segmentation masks produced via atlas-registration to motion-corrupted 2D stacks of short-axis cardiac MR images \citep{Duan2019}, which improved topological accuracy of 3D FCN predictions. Another approach involved simultaneous training of parallel network branches for registration and segmentation, providing a form of regularization on each branch \citep{Xu2019}. Loss terms which explicitly penalize topologically undesirable predictions using persistent homology \citep{Clough2019, Byrne2020} have also been proposed, improving topological accuracy over pixel-wise segmentation models. Incorporation of point cloud prediction as an intermediate representation in a 3D segmentation network has also demonstrated improvements in topological consistency and segmentation accuracy \citep{ye2020}. Finally, prediction of signed distance fields defining segmentation boundaries has also been proposed to improve segmentation accuracy \citep{Li2020,Tilborghs2020}. While each of these approaches have their inherent advantages, they do not guarantee a target topology at test time. 

Another class of deep learning models which has received growing attention utilises a PCA shape model of surface meshes of training cases, the weights of which can be predicted directly by a CNN for a target image \citep{Milletari2017, Bhalodia2018, Tothova2018, Tothova2020, Adams2020}. These models have been proposed to predict 3D surface meshes both from 3D images \citep{Bhalodia2018, Adams2020} and in settings where only sparse 2D images are available \citep{Milletari2017, Tothova2018, Tothova2020}. A hierarchical variational auto-encoder has also been proposed for the latter setting \citep{Cerrolaza2018}, where shape parameters are implicitly encoded in the latent variables. While such models directly encode a topologically consistent structure, as with classic PCA shape models they potentially suffer from over-constraining shape descriptors to the training set and may not be sensitive to subtle anatomical variations or out-of-distribution examples at test time. While these models show promise, with the added benefit of uncertainty quantification \citep{Tothova2020, Adams2020}, such models have not been benchmarked against state-of-the-art 3D semantic segmentation models in the setting where dense 3D images are available.

Recently, there has also been a growing interest in the field of deep learning for image registration \citep{Haskins2020}, following the seminal work on spatial transformer networks (STN) \citep{Jaderberg2015}, with a number of models proposing to use registration to propagate labelmaps to target images as a means of segmentation \citep{Lee2019b,Balakrishnan2018,Dalca2019,Dong2020,Mansilla2020}. A prominent method in the recent literature is the VoxelMorph framework for unsupervised and semi-supervised learning of image registration \citep{Balakrishnan2018}. This framework uses an encoder-decoder type CNN to predict a dense displacement field used to register a source image to a target image, which can also be used to propagate a source labelmap to the target image. The model can be trained in a fully unsupervised setting with loss terms including image similarity, and penalties on deformation field smoothness and magnitude. Auxiliary losses that consider labelmap overlap have also been used for semi-supervised learning of image registration \citep{Balakrishnan2018,lee2020labeled}, where a subset of the training data may have labelmap annotations. VoxelMorph proved highly effective for test time image registration when trained on a large set of brain MR images, with the observation that instance-specific optimization of the predicted displacements at test time produced improved performance, particularly when fewer training examples were available \citep{Balakrishnan2018}. \cite{Mansilla2020} recently proposed AC-RegNet, an image registration model regularized with a shape-aware auto-encoder conditioned on labelmaps during training, which yielded more anatomically plausible predicted displacement fields at test time for 2D lung X-ray images compared to a pixel-wise baseline. They also demonstrated that AC-RegNet could be used for multi-atlas segmentation. \cite{Dong2020} proposed a deep learning framework with adversarial consistency for registration of a pre-defined population-derived atlas image and labelmap to a target image. Adversarial image and labelmap pairs were used to encourage the model to predict more accurate deformations, and both affine and non-rigid transformations were predicted by separately trained CNNs. The model was trained and evaluated using a limited set of 3D echocardiography images with annotations of the left ventricular myocardium. The performance of the proposed method improved over state-of-the-art voxel-wise segmentation methods, although only 25 cases were used for training with limited data augmentation (only rotations around the $z$-axis), which provides a sub-optimal setting for training a voxel-wise segmentation model. \cite{Dalca2019} proposed a framework that directly parameterizes a template (or atlas) image volume to be jointly learned with a registration model that registers the template image to target images. The learnable template could also be conditioned on parameters of interest, such as age and sex, and the method was evaluated with a dataset of brain MR images. The template image can be likened to a statistical atlas image learned from the training set. The authors demonstrated that a template \emph{labelmap} could be constructed \emph{after} training by registering training images to the constructed atlas \emph{image} and propagating their labelmaps and subsequently fusing them. Image segmentation at test time was then performed by propagating the constructed template labelmap to target images, producing promising results compared to VoxelMorph \citep{Balakrishnan2018}. 

Finally, the image-and-spatial-transformer network (ISTN) was proposed in \citep{Lee2019a}, where an image transformer network (ITN) is trained jointly with a spatial transformer network (STN). This approach proposed to generate intermediate representations of SoI from an input source and target image pair with an ITN, which are passed downstream to the STN to register the inputs. Similarly to VoxelMorph, loss terms optionally included image similarity, deformation field penalties and terms which leveraged ground-truth labelmaps for SoI-guided registration. A demonstrated advantage of ISTN is that the STN parameters can be optimized on specific instances at test time to register ITN predictions of a source and target image pair, which improves overlap of the SoI compared to both a learned registration of the images via their intermediate representation (i.e. a single pass of the STN at test time) as well as a model that registered images without intermediate representations, such as Voxelmorph \citep{Balakrishnan2018}.

\begin{figure}[!h]
    \centering
    \includegraphics[width=\linewidth]{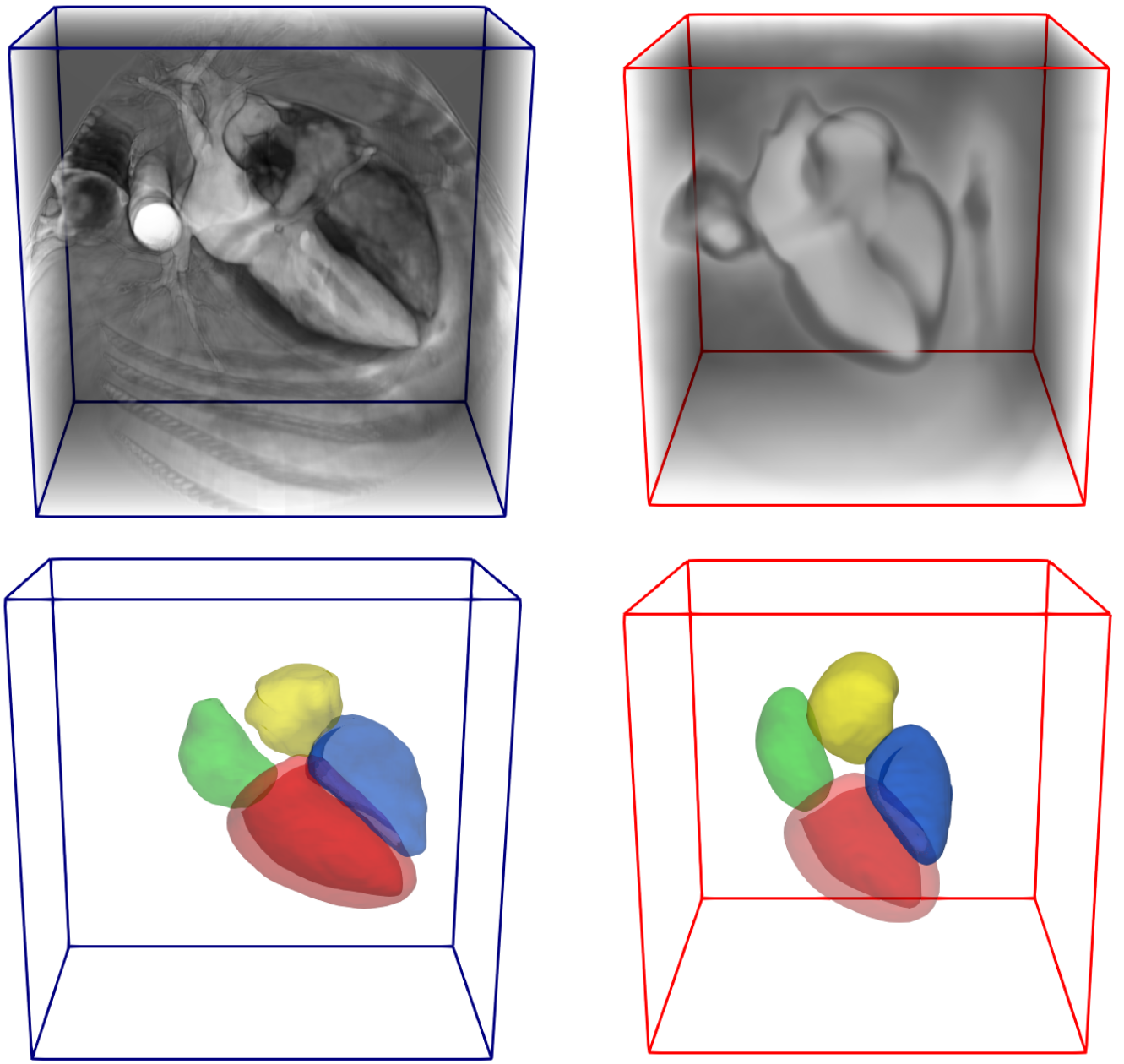}
    \caption{Intensity projections of images (top) and labelmaps (bottom) of a CCTA training case (left) and the constructed atlas (right). Structures depicted include the left ventricle myocardium (red), right ventricle blood pool (blue), right atrial blood pool (yellow) and left atrial blood pool (green).}
    \label{fig:volumes}
\end{figure}

\paragraph{Contributions}
Atlas-ISTN extends the ISTN framework and other proposed works to jointly learn a population-derived atlas together with a model which performs segmentation and registration. The contributions of the Atlas-ISTN framework are:

\begin{enumerate}
\item An all-in-one deep learning framework for atlas construction, registration and segmentation;
\item A robust segmentation system which improves over baseline segmentation and registration models;
\item A method for construction of an unbiased population-derived atlas within a deep learning framework;
\item Topological guarantees on test time segmentations via registration of a constructed atlas labelmap;
\item A deep learning framework that provides inter-subject correspondences of SoI via a mapping to atlas space.
\end{enumerate}

\section{Methods}

For many segmentation tasks, the topology of the target SoI is known \emph{a priori}. For example, in medical image analysis, the chambers of the heart typically conform to the same spatial arrangement from subject to subject, with variation in shape, size, orientation and wall thickness. Voxel-wise segmentation models do not take advantage of this \emph{a priori} knowledge, and can produce spurious segmentations or topologically inconsistent predictions, particularly with limited training data. The proposed Atlas-ISTN framework seeks to address this issue by both learning a voxel-wise prediction and learning to fit a topologically consistent atlas labelmap to the SoI, reducing noisy predictions and ensuring topological plausibility.  

The main components of the framework consist of the joint training of the ITN and STN (Sec. \ref{subsec:network}), the simultaneous construction of the atlas (Sec. \ref{subsec:atlas}), and test time refinement of the atlas (Sec. \ref{subsec:refinement}). These components are put into context below, and described in detail in the following sections. 

Similarly to ISTN \citep{Lee2019a}, the Atlas-ISTN architecture consists of two sequential blocks: an ITN and a STN. The purpose of the ITN is to learn an intermediate representation from an input image which is useful to a downstream task, for example registration of the SoI \citep{Lee2019a}. One option for this intermediate representation is a semantic segmentation, which requires labelled data during training. Such an intermediate representation can also be leveraged at test time for further refinement of the STN parameters. 

The STN in the Atlas-ISTN framework learns a spatial transformation to map the ITN prediction to an atlas labelmap and vice versa, in order to maximize the agreement between a ground-truth labelmap and the deformed atlas labelmap (and vice versa). The ITN and STN are described in Section \ref{subsec:network}, and the overall framework is shown in Fig. \ref{fig:framework}. 

Instead of registering pairs of images or intermediate representations as done in many recent works \citep{Lee2019a,Balakrishnan2018,Dalca2018a,Mansilla2020}, we propose to register predicted labelmaps to an atlas labelmap which is constructed during training. The most closely related recent works include \citep{Dalca2019} where an atlas image of the brain is learned during training, optimizing alignment of this constructed atlas image with training images. In \citep{Dong2020}, group-wise registration was first used to generate an atlas image and labelmap that were then fixed during training. In this work, we introduce a method to construct a multi-structure atlas labelmap and image on-the-fly during training, which is generated in tandem to optimizing the parameters of the ITN and STN. This approach draws on ideas from classic atlas construction methods, where images are registered to a reference space and averaged in an iterative procedure \citep{Guimond2000,joshi2004unbiased}. The proposed approach is described in Section \ref{subsec:atlas}. 

Finally, the Atlas-ISTN framework enables test time refinement in a similar way to the ISTN framework \citep{Lee2019a}, but instead of registering the segmentations predicted by the ITN for a pair of input images, the STN parameters can be fine-tuned to maximize the agreement between the ITN prediction of an input image and the deformed atlas labelmap. This helps overcome limitations of the first-pass of a learned registration model at test time, particularly for out-of-distribution cases which often result in poor alignment of SoI. With appropriate choice of test time regularization, refinement allows for improved registration of the atlas labelmap with the SoI predicted by the ITN, circumventing spurious ITN predictions and false negatives. This process is described further in Section \ref{subsec:refinement}.   

In the following sections, we present the transformation model (Sec. \ref{subsec:transformations}) neural network architecture (Sec. \ref{subsec:network}), the method by which the atlas is learned (Sec. \ref{subsec:atlas}), and the refinement procedure which can be followed at test time to optimize the STN to improve the fit of the constructed atlas to the ITN prediction (Sec. \ref{subsec:refinement}).

\begin{figure*}[!ht]
    \centering
    \includegraphics[width=0.85\linewidth]{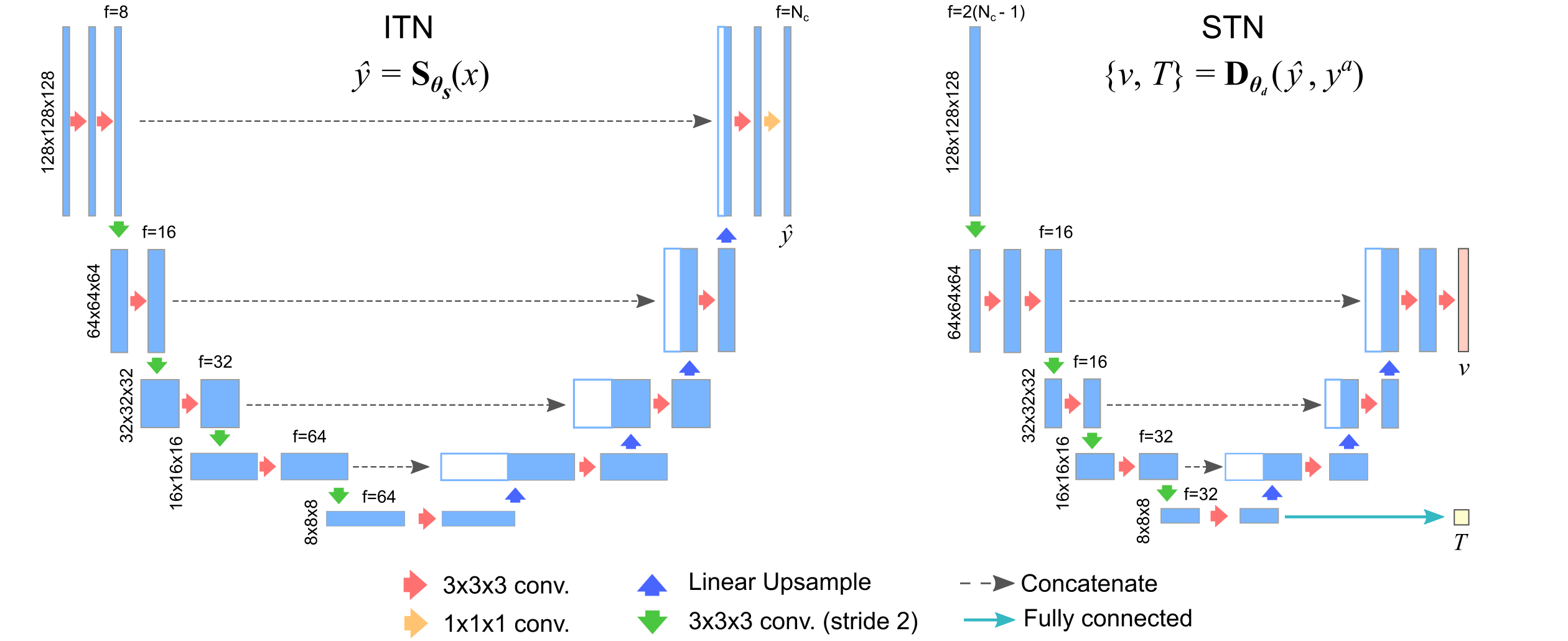}
    \caption{An overview of the (3D) ITN and STN architectures. Note, $3\times3\times3$ conv. layers (red and green arrows) are followed by a ReLU activation, except for in the final layer of the STN. Outputs of the STN, $v$ and $T$, are passed to the Transformation Computation Module in Fig. \ref{fig:compute_deformations}.}
    \label{fig:architectures}
\end{figure*}

\subsection{Deformations}\label{subsec:transformations}

In this work we model both affine and non-rigid deformations, which are commonly used in two-stage atlas registration methods particularly for cardiac image data \citep{Lamata2014,Bai2015}. We consider the following composition of an affine transformation:

\begin{equation}
    T = M_t R_{\theta} D_s,
\end{equation}

where $M_t$, $R_{\theta}$ and $ D_s$ are the translation, rotation and scaling matrices (we exclude shearing in this work). This affine transformation provides a coarse registration between a source and target image. Affine registration is often used as a pre-alignment step \citep{Bai2015,Dong2020}, providing a more optimal starting point for a non-rigid registration to be performed. This type of pre-alignment is also important for applications with brain images, so much so that it is built into standardized processing pipelines, and thus most recent work exploring learning deformations for brain image data has only utilized a non-rigid component in their registration models \citep{Dalca2018a,Balakrishnan2018,Krebs2019}. While typically used as a pre-alignment, an affine transformation can also be optimized jointly with a non-rigid transformation \citep{Stergios2018}, where the affine component would be expected to account for large-scale deformations and coarse alignment of source and target images. In addition, by explicitly modeling the global transformation, a rigid normalization of pose is not penalized by the regularity commonly imposed on the local deformations.

As non-rigid component, we choose a diffeomorphic transformation model parameterized by a stationary velocity field (SVF), in order to guarantee preservation of the topology of the constructed atlas after deformation, and to ensure invertibility \citep{Arsigny2006, Ashburner2007}. The differential equation describing the evolution of a deformation generated by a SVF denoted by $v$, is given by

\begin{equation}
 \frac{\text{d}\phi}{\text{d}t} = v \large{(} \phi^{(t)} \large{)}.
\end{equation}

The deformation at $t=1$ ($\phi^{(1)}$) is obtained by integration of this ordinary differential equation (ODE) over unit time starting with the identity $\phi^{(0)}(x) = x$ \citep{Ashburner2007}. Subsequently, we denote this deformation as $\phi$ to simplify notation. In the theory of Lie groups, solving this ODE is equivalent to computing the exponential map of the flow field $v$ (a member of the \emph{Lie algebra}), i.e.,

\begin{equation}\label{eq:exp_v}
 \phi = \text{exp}(v).
\end{equation}

The inverse transformation $\phi^{-1}$ can thus be obtained by the exponentiation of the negative SVF. In practice, the exponential map is computed efficiently via scaling and squaring \citep{Arsigny2006, Ashburner2007}.

\subsection{Atlas-ISTN Model}\label{subsec:network}

\paragraph{Image Transformer Network} Given input images and ground-truth labelmaps, $X = \{\,x_i, y_i\,\}$, the ITN learns the mapping

\begin{equation}\label{eq:itn_forward_pass}
\hat{y}_i = \mathrm{\mathbf{S}}_{\theta_s}(x_i),    
\end{equation}

where $\hat{y}_i$ are the (multi-channel) logits of a labelmap prediction. Both $\hat{y}_i$ and $y_i$ contain a background channel and as many foreground channels as structures in the training data. 

The ITN can be any suitable pixel-wise segmentation model. In this work 2D and 3D U-net models are used \citep{Ronneberger2015, Cicek2016}, which consist of convolutional layers in an encoder-decoder format with skip connections between the corresponding spatial scales of the encoder and decoder (Fig. \ref{fig:architectures}, left). Coarser spatial scales in the encoder are achieved via strided convolutions, and in the decoder bilinear up-sampling is used\footnote{Improved performance was found over transposed convolutions.}.  

\begin{figure}[!hb]
    \centering
    \includegraphics[width=6cm]{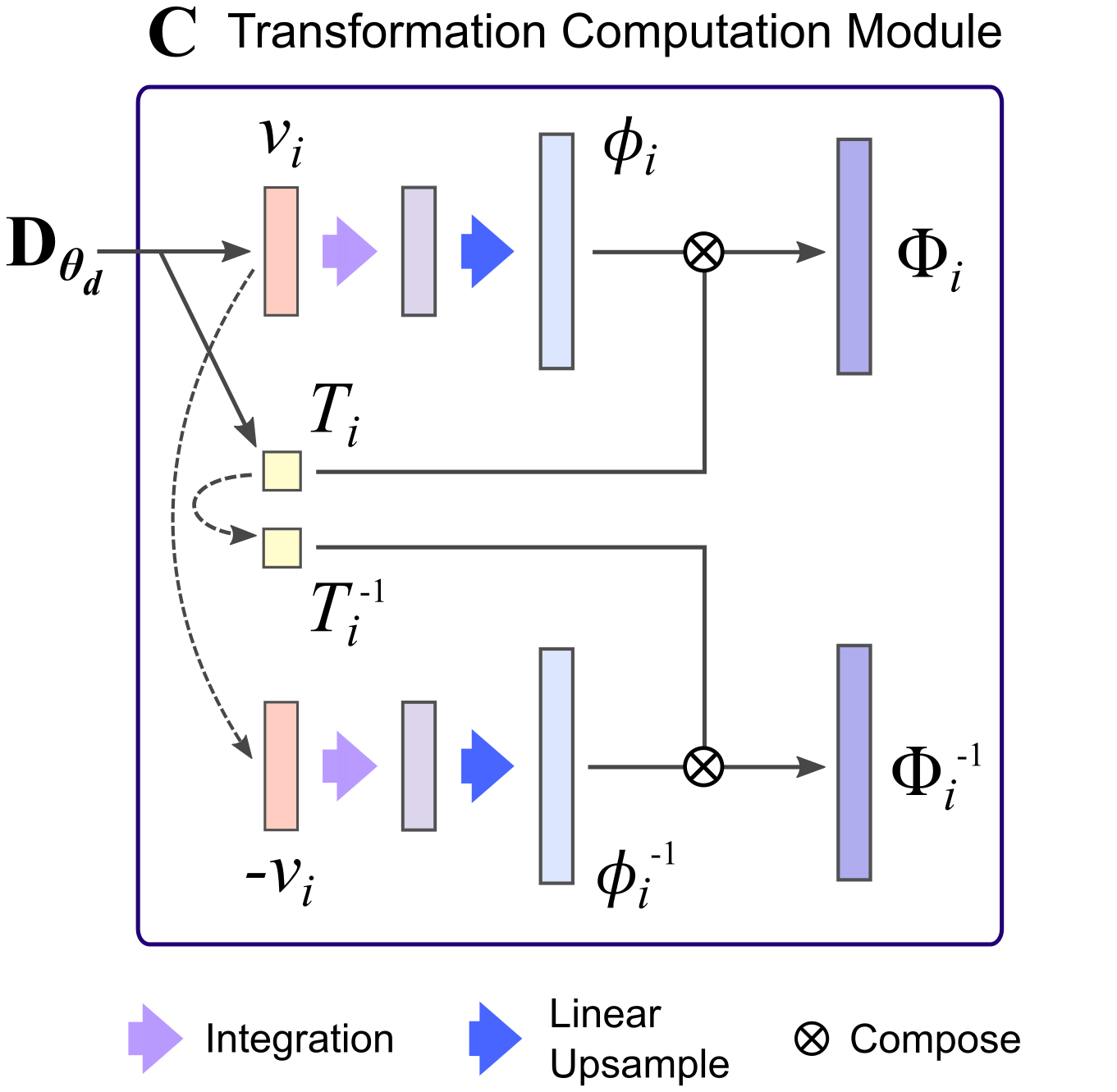}
    \caption{The Transformation Computation Module. The predicted SVF ($v$) is integrated via scaling and squaring, before linear upsampling. The resulting transformation ($\phi$) is composed with the predicted affine parameters ($T_i$). Computation of the inverse and forward transformations are shown at the top and bottom, respectively. Compositions are defined in Eqs. \eqref{eq:forward_def} and \eqref{eq:backward_def}.}
    \label{fig:compute_deformations}
\end{figure}

\paragraph{Spatial Transformer Network} We propose an STN which learns the mapping

\begin{equation}\label{eq:stn_forward_pass}
\{\,v_i, \, T_i\,\} = \mathrm{\mathbf{D}}_{\theta_{d}}(\hat{y}_i, y^a),
\end{equation}

where the concatenation of (1) the ITN logits $\hat{y}_i$ and (2) the atlas labelmap $y^a$ make up the input tensor. The STN produces two outputs including (1) a SVF, $v_i$, and (2) an affine transformation matrix, $T_i$, which are subsequently processed and composed into a final deformation field (see Fig. \ref{fig:compute_deformations}) by the function 

\begin{equation}\label{eq:compute_deformations}
    \{\,\Phi_i, \, \Phi_i^{-1}\,\} = \mathrm{\mathbf{C}}(v_i, T_i).
\end{equation}

Similarly to \citep{Stergios2018}, joint prediction and composition of affine and non-rigid deformations by the STN is performed. $T_i$ is represented as a $3\times4$ matrix, and $v_i$ has a size of $3 \times \frac{N_x}{2} \times \frac{N_y}{2} \times \frac{N_z}{2}$ (in the 3D setting), where $N_i$ denotes the input image size in each spatial dimension. Modeling both affine and non-rigid deformations provides greater flexibility\footnote{See the next section on \emph{Losses}.} to generate transformations which capture the inherent spatial differences between SoI in many types of patient scans where images are not traditionally pre-aligned for a target anatomical structure, such as with CCTA images. 

Within the Transformation Computation Module, the forward and inverse transformations are the compositions:

\begin{equation}\label{eq:forward_def}
    \Phi_i \,\,\, = T_i \circ \phi_i,
\end{equation}
\begin{equation}\label{eq:backward_def}
    \Phi_i^{-1} = \phi^{-1}_i \circ T_i^{-1}.
\end{equation}

The deformations are applied to each labelmap channel, $j$, independently, such that a transformed labelmap is obtained by $y_{i,j} \circ \Phi_i$. A forward pass through the network via ITN (Eq. \eqref{eq:itn_forward_pass}), STN (Eq. \eqref{eq:stn_forward_pass}) and Transformation Computation Module (Eq. \eqref{eq:compute_deformations}) can be expressed concisely as:

\begin{equation}\label{eq:forward_pass}
\{\, \Phi_i, \, \Phi^{-1}_i \,\} = \mathrm{\mathbf{C}}\Big(\mathrm{\mathbf{D}}_{\theta_d}\big(\mathrm{\mathbf{S}}_{\theta_s} (x_i), y^a\big)\Big).  
\end{equation}

\paragraph{Losses}\label{par:losses}

A mean squared error (MSE) loss is used for the supervised learning of the ITN weights ($\theta_s$), i.e.,

\begin{equation}\label{eq:seg_loss}
    L_s = \frac{1}{n} \sum_i^n \|y_{i} - \hat{y}_{i}\|^2,
\end{equation}

where $i$ is the image index and $n$ the total number of images. In experiments where an ITN was trained on its own with different losses (not reported in results), similar performance was obtained with a MSE loss compared to using cross-entropy. A MSE loss for $L_s$ was found to produce the best convergence and overall results for Atlas-ISTN\footnote{Careful weighting of loss terms was required when training with cross-entropy for $L_s$ while using MSE for $L_{s2a}$ and $L_{a2s}$, which still under-performed compared to using an MSE loss for all three terms.}. Two additional MSE losses, the atlas-to-segmentation loss ($L_{a2s}$) and the segmentation-to-atlas loss ($L_{s2a}$) are used:

\begin{equation}\label{eq:a2s}
    L_{a2s} = \frac{1}{n} \sum_{i=0}^n \sum_{j=1}^c \|y_{i, j} - y^a_j\circ \Phi^{-1}_i\|^2,
\end{equation}

\begin{equation}\label{eq:s2a}
    L_{s2a} = \frac{1}{n} \sum_{i=0}^n \sum_{j=1}^c \|y_{i, j}\circ \Phi_i\ - y^a_j\|^2,
\end{equation}

where $j$ denotes the labelmap channel index (and $j=0$ corresponds to the background channel), $i$ denotes the case index, and $n$ the number of training images. $L_{a2s}$ encourages accurate transformation of the atlas labelmap ($y^a$) to the ground-truth labelmap ($y_i$), despite potential noise in the ITN prediction ($\hat{y}_i$) which is an input to the STN. $L_{s2a}$ encourages accurate transformation of the ground-truth labelmap ($y_i$) to the atlas labelmap ($y^a$), which aides the atlas learning process (in Section \ref{subsec:atlas}).      

A regularization term is also used to encourage smoothness of the predicted non-rigid deformation fields, i.e.,

\begin{equation}
L_{reg} = \sum_i^n \| \nabla \phi_i \|^2.    
\end{equation}

The overall loss is

\begin{equation}\label{eq:training_loss}
    L = L_s + \omega \, (L_{a2s} + L_{s2a} + \lambda L_{reg}),
\end{equation}

where $\lambda$ controls the effect of the regularization term, and $\omega$ controls the overall weighting of the deformation-related loss terms. The segmentation loss ($L_s$) provides gradients only for updating the ITN weights ($\theta_s$) while all other loss terms also contribute to gradients which update the STN weights ($\theta_d$). Note that since $L_{reg}$ only penalizes the non-rigid component of the deformation ($\phi$), this encourages the affine component of the STN prediction ($T$) to learn, and normalize for, global pose and scale, in turn encouraging the SVF to focus on learning local deformations.

\paragraph{Other Practicalities} The SVF is predicted at half the resolution of the input image, which was found to produce smoother deformation fields and better convergence during training, as well as reduced memory overhead. The full resolution deformation fields (as used in Eqs. \eqref{eq:forward_def} and \eqref{eq:backward_def}) are obtained by linear upsampling, shown schematically in Fig. \ref{fig:compute_deformations}. 

Architecturally, the STN is similar to the ITN in terms of number of spatial scales, the encoder-decoder structure and skip connections, as depicted in Fig. \ref{fig:architectures}. In both components, batch normalization was found to perform worse and was therefore not used. Trilinear upsampling was used rather than transposed convolutions for upsampling throughout the network, which removed checkerboard artifacts and improved overall performance. 

In practice, the foreground channels of $\hat{y}_i$ and $y^a$ are concatenated and passed as inputs to the STN, as we found removing the background channel improved test time performance, improved training stability and reduced memory overhead. The number of input channels to the STN therefore is equivalent to $2(N_c - 1)$, where $N_c$ is the number of channels (including background) of the training labelmaps. Having multiple input channels for the STN (compared to just 1 for the ITN) led to the use of more filters in the first convolutional layer of the STN, which was found to improve convergence of the STN during training. To reduce memory overhead associated with this increase in filters, the first convolutional layer of the STN has a stride of 2, immediately leading to a spatially reduced feature map. This did not negatively impact model performance, likely due to the fact that $v_i$ is predicted at half the resolution of the input images, where the SVF-derived deformation fields ($\phi_i$ and $\phi_i^{-1}$) are linearly upsampled to the input image resolution.

\subsection{Atlas Construction}\label{subsec:atlas}

Unlike in \citep{Dalca2019} where a volumetric atlas of the training \emph{images} is learned over the course of model training, in this work we propose a method to jointly construct volumetric atlases of the training images and \emph{labelmaps}. This is achieved without explicitly parameterizing atlas voxels with learnable weights as in \citep{Dalca2019}, but rather we draw inspiration from classic atlas construction work \citep{Guimond2000,joshi2004unbiased}. The ITN and the STN are trained jointly, and the atlas labelmap ($y^a$) and image ($x^a$) are updated at the end of each epoch by warping training data to atlas space via a forward pass of Atlas-ISTN and averaging across all samples. The update procedure for both atlas labelmap and image are the same, so we denote both images ($x$) and labelmaps ($y$) by $z$ and the respective atlas by $z^a$ in the following. The atlas is initialized and updated by

\begin{equation}\label{eq:atlas_update}
 z^a_{j,\,t} =
 \begin{cases}
    \frac{1}{n} \, \sum_{i=0}^{n-1} (z_{i,\,j})\,, & \,\,\, t=0 \\
    (1 - \eta) \, z_{j,\,t-1}^a + \eta \, \tilde{z}_{j,\,t}^a\,, & \,\,\, t \ge 1
 \end{cases}
\end{equation}

where $i$ denotes the case index, $j$ the channel index, $t$ the epoch, and $n$ the total number of training cases. The atlas is initialized (at $t=0$) as the mean of all the (undeformed) training cases. The rate at which the atlas is updated is determined by $\eta$, and $\tilde{z}_t^a$ is the mean of the transformed training cases: 

\begin{equation}\label{eq:atlas_mean}
 \tilde{z}^a_{j,\,t} =  \frac{1}{n}\sum_{i=0}^{n} (z_{i,j} \circ \Phi_{i,\,t}), \quad\quad\quad\quad \,\,\, t \ge 1
\end{equation}

 At the end of each training epoch, a forward pass through the network (Eq. \eqref{eq:forward_pass}) is used to warp each training case to atlas space, from which the channel-wise mean atlas is computed (Eq. \eqref{eq:atlas_mean}). The updated atlas labelmap at the end of each epoch, ($t \ge 1$) in Eq. \eqref{eq:atlas_update}, is then used during the following epoch in the computation of the losses $L_{a2s}$ (Eq. \eqref{eq:a2s}) and $L_{s2a}$ (Eq. \eqref{eq:s2a}). The atlas labelmap and image are shown in Fig. \ref{fig:volumes} alongside the labelmap and image of a randomly selected training case from the CCTA dataset.  

For the proposed Atlas-ISTN model, we focus on the use of the atlas labelmap $y^a$ to improve segmentation performance at test-time, described in the next section. The benefits of learning an atlas labelmap are that (1) an unbiased labelmap is produced through the training process, as opposed to using a fixed atlas, and (2) the final STN is optimized to warp this unbiased atlas labelmap to the ITN prediction, which positions the STN weights in a setting well-suited for test time refinement (as opposed to optimizing STN weights for test-time refinement from scratch). Creating an unbiased atlas via established methods such as group-wise registration \citep{joshi2004unbiased} as an alternative to (1) could also be done, but is not required since Atlas-ISTN is able to construct the atlas during training. 

\subsection{Test Time Refinement}\label{subsec:refinement}

At test time, an observation made in \citep{Lee2019a} was that the overlap of source and target image SoI could be further improved with test time optimization of the STN weights to register the source and target ITN predictions. In this work, the STN weights are optimized for the registration of the constructed atlas labelmap to the ITN prediction of a target image, referred to throughout the text as refinement. For a given target image, $L_{a2s}$ (Eq. \eqref{eq:a2s}) and $L_{s2a}$ (Eq. \eqref{eq:s2a}) can be repurposed to optimize the alignment between the atlas labelmap ($y^a$) and the ITN logits ($\hat{y}$) instead of training labels ($y$) used during training,

\begin{equation}\label{eq:a2s_ref}
    L^{*}_{a2s} = \sum_{j=1}^c \|\hat{y}_{i, j} - y^a_j\circ \Phi^{-1}_i\|^2,
\end{equation}

\begin{equation}\label{eq:s2a_ref}
    L^{*}_{s2a} = \sum_{j=1}^c \|\hat{y}_{i, j}\circ \Phi_i\ - y^a_j\|^2,
\end{equation}

where the overall refinement loss is given by:

\begin{equation}\label{eq:refine_loss}
    L^* = \beta^{*} L^*_{a2s} + \gamma^* L^{*}_{s2a} + \lambda^* L_{reg},
\end{equation}

with $\beta^*$, $\gamma^*$ and $\lambda^*$ corresponding to weightings on the atlas-to-segmentation, segmentation-to-atlas and regularization terms, respectively. During refinement, ITN weights are fixed and only STN weights are updated. In the presence of noise in $\hat{y}_i$, larger values of $\lambda$ can encourage a more rigid registration that retains more of the atlas shape, and in turn circumvent spurious segmentations or holes in $\hat{y}_i$. In practice we find that refinement without the segmentation-to-atlas loss (i.e. setting $\gamma^*=0$) slightly improves the accuracy of the final deformed atlas while reducing memory overhead. The deformed atlas after refinement is given by:

\begin{equation}\label{eq:refine_transform_atlas}
y^{a*}_i = y^a \circ \Phi^{*-1}_i,
\end{equation}

where for a given input image $x_i$, $\Phi^{*-1}_i$ is obtained from a forward pass of Atlas-ISTN after refinement,
 
\begin{equation}\label{eq:refine_forward_pass}
 \{ \, \Phi^{*}_i, \Phi^{*-1}_i \, \} = \mathrm{\mathbf{C}}\Big(\mathrm{\mathbf{D}}^*_{\theta_d}\big(\mathrm{\mathbf{S}}_{\theta_s} (x_i), y^a\big)\Big),  
\end{equation}

where $\mathrm{\mathbf{D}}^*_{\theta_d}$ denotes the STN with refined weights after $K$ refinement iterations minimizing the loss in Eq. \eqref{eq:refine_loss} via stochastic gradient descent. This process is illustrated in the lower panel of Fig. \ref{fig:framework}.  

\section{Experiments and Results}\label{sec:experiments}

\subsection{Letter B}\label{subsec:letter_b}

To illustrate some of the key properties of Atlas-ISTN for image segmentation, let us initially consider a simple toy example based on a binary image of the letter B. This example acts as proof-of-concept and allows us to demonstrate the behaviour of Atlas-ISTN in a controlled setting. We show some qualitative and quantitative results in the following.

\begin{figure}[!h]
    \centering
    \includegraphics[width=\linewidth]{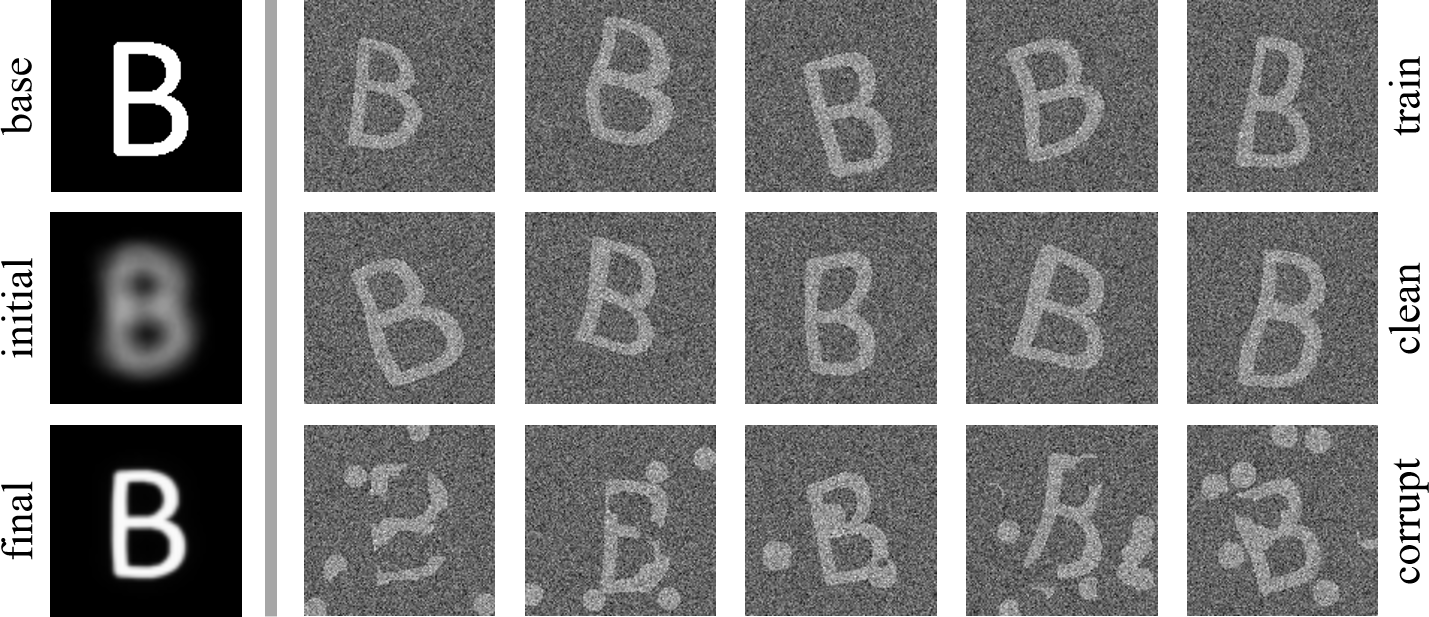}
    \caption{The left-most column illustrates the base letter B (top), the initial atlas labelmap at the start of training defined in Eq. \eqref{eq:atlas_mean} at $t=0$ (middle), and the recovered letter B atlas labelmap after training (bottom). The other images are examples from the training set (top row) and clean (middle row) and corrupt test set (bottom row) generated randomly from the base letter B.} 
    \label{fig:synth2d-overview}
\end{figure}

\paragraph{Data Description}
Starting from a single, binary 2D image of the letter B, we generate warped instances of this base image using random affine transformations composed with random B-spline deformations. For each warped binary image we generate a corresponding intensity image with additive Gaussian noise. We use 1,000 of these image and labelmap pairs for training an Atlas-ISTN. We further use a hold-out set of 100 pairs to test the Atlas-ISTN, once on a clean version of the test set (coming from the same distribution as the training data) and a corrupted version (with random clutter added to the intensity images). Visual examples of the training set and the clean and corrupted test sets together with base letter B, the initial and resulting constructed atlases are shown in Fig~\ref{fig:synth2d-overview}. This synthetic 2D data is provided together with our source code such that the following results can be fully replicated.

\begin{figure*}[!h]
    \centering
    \includegraphics[width=\linewidth]{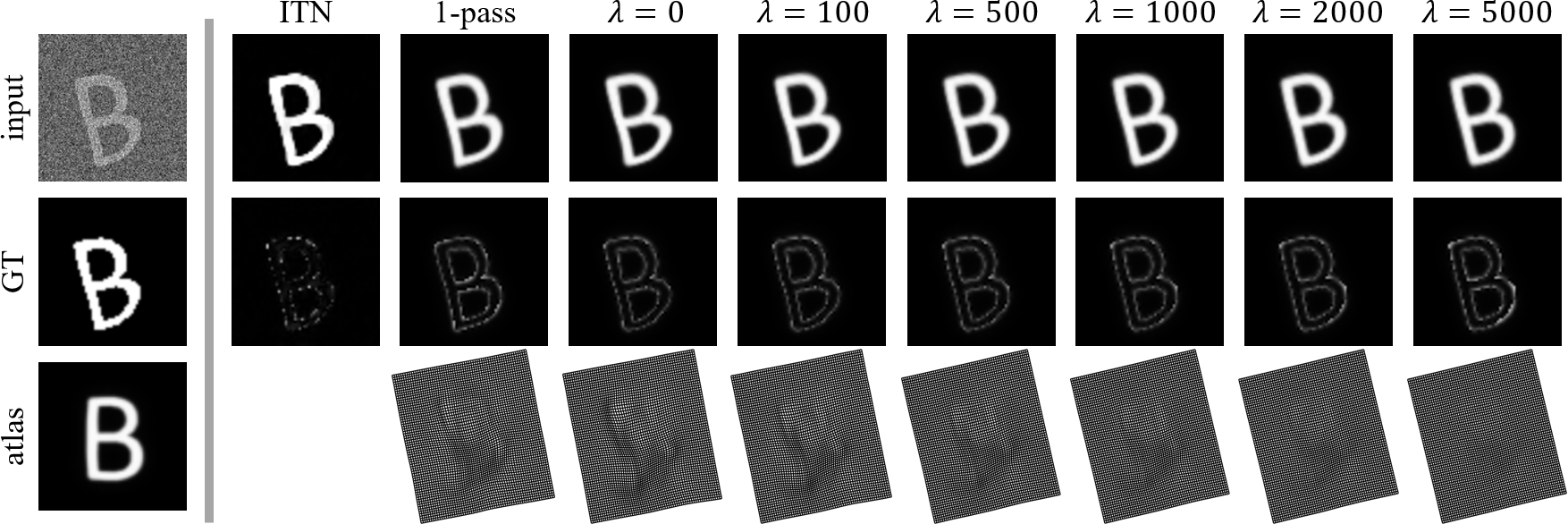}
    \caption{Qualitative results for the 2D toy data with test data coming from the same distribution as the training data. Both the ITN and 1-pass Atlas-ISTN yield accurate segmentations. test time refinement with increasing regularization weight $\lambda$ affects the smoothness of the final transformation.}
    \label{fig:synth2d-results-clean}
\end{figure*}

\begin{figure*}[!h]
    \centering
    \includegraphics[width=\linewidth]{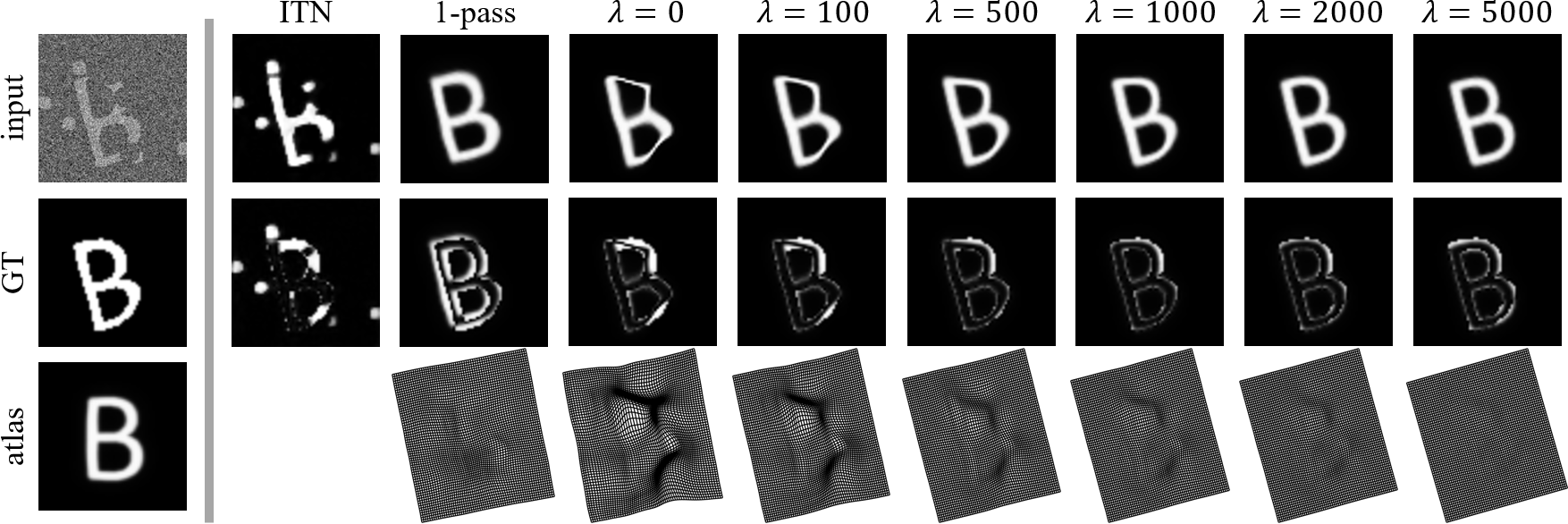}
    \caption{Qualitative results for the 2D toy data with corrupted, out-of-distribution test data. The ITN yields many false positives and false negatives and is topologically implausible. The 1-pass Atlas-ISTN yields a reasonable atlas alignment despite the corrupted input data. test time refinement with increasing regularization weight $\lambda$ can yield accurate and topologically plausible segmentations.}
    \label{fig:synth2d-results-noisy}
\end{figure*}

\paragraph{Qualitative Results}
In the case of clean test data, we make the following observations: As expected, the ITN provides nearly perfect segmentations of the input images, and equally the 1-pass Atlas-ISTN predicts an accurate alignment of the constructed atlas to the ground truth. We run test time refinement with six different regularization weights and observe the effect of increased regularization on the final transformation. This example confirms that when sufficient training data is available, and the test data comes from the same distribution, the 1-pass Atlas-ISTN is en-par with an ITN-based segmentation in terms of accuracy with the added benefit that the resulting segmentations of the Atlas-ISTN come with correspondences to the constructed atlas space. Test time refinement further allows to control the degree of deformation through the regularization weight, but may be considered optional as it may not add significant improvements in segmentation accuracy.

In reality, however, test data often does not come from the exact same distribution as the training data and this may negatively affect the predictive performance of a trained network. Test images, for example, might exhibit variations due to artifacts or pathology which were not captured in the training data. This is simulated here by testing the above Atlas-ISTN on a corrupted version of the test set. We observe that the ITN now fails to accurately segment the structures of interest yielding both many false positives and false negatives. Still, the predicted segmentation may provide useful information for subsequent refinement in our Atlas-ISTN framework. The 1-pass prediction of the Atlas-ISTN is also affected by the noisy ITN output yielding a sub-optimal alignment of the atlas, yet providing a good initial atlas alignment. Here, the benefits of the test time refinement become clear. This test-specific optimization of the STN network weights results in plausible segmentations of the corrupted test images, removing both false positives and negatives and adhering to the topology of the constructed atlas. Again, we observe the effect of the regularization weight which provides control over how closely we wish to stay to the constructed atlas up to affine transformations.

\begin{figure}[!h]
    \centering
    \includegraphics[width=\linewidth]{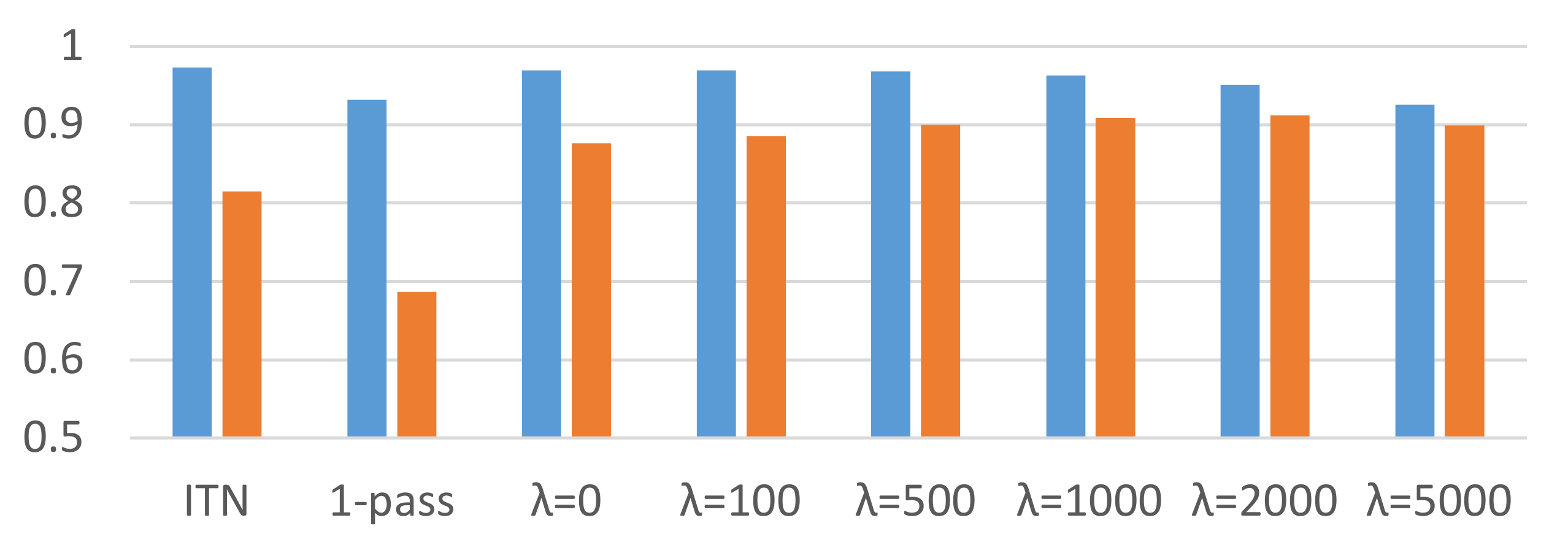}
    \includegraphics[width=\linewidth]{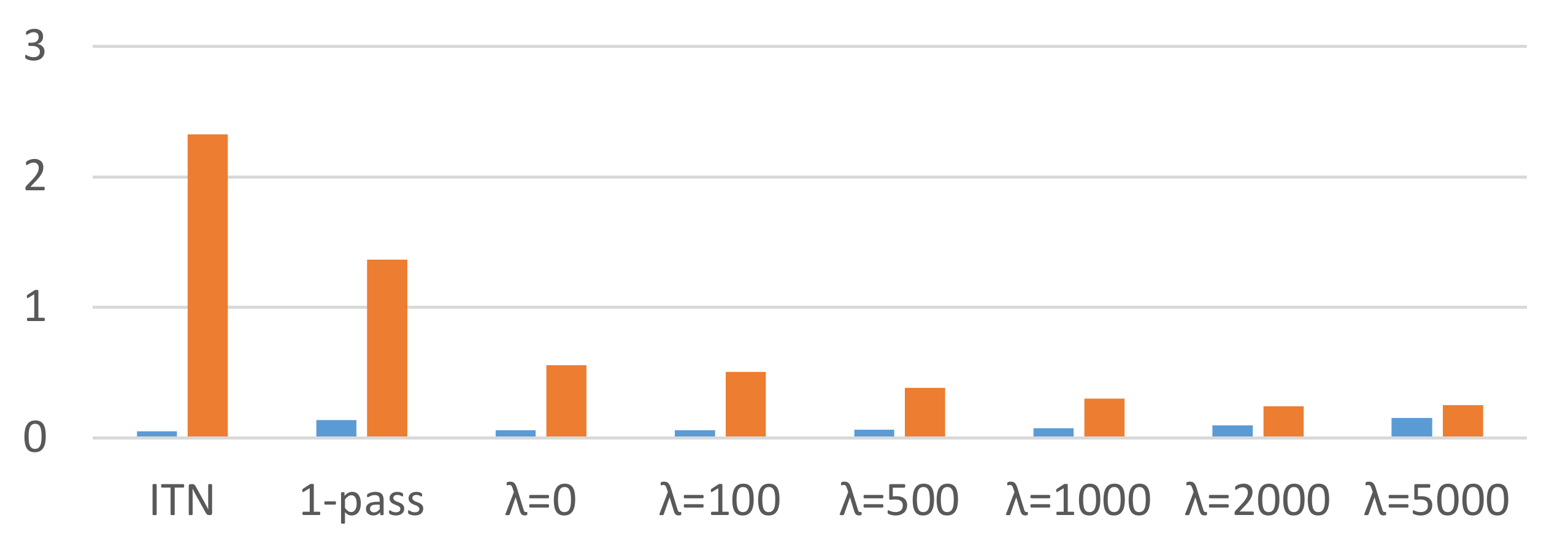}
    \includegraphics[width=\linewidth]{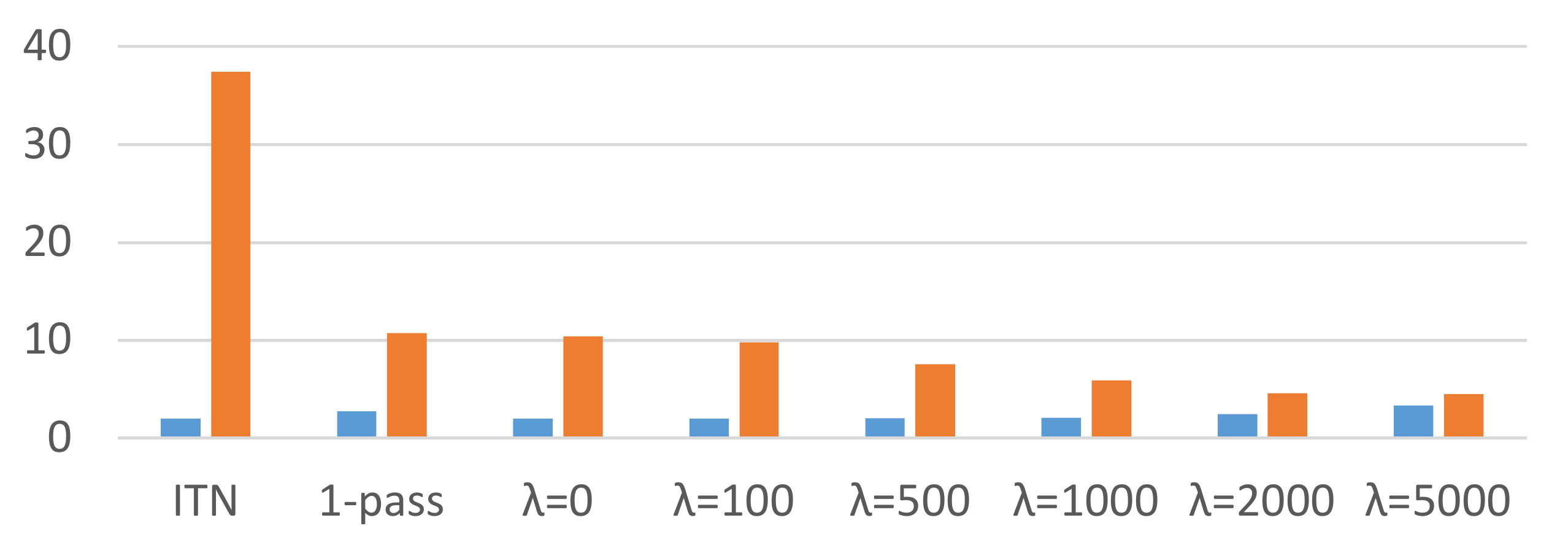}
    \caption{Quantitative results for the 2D toy data with test data from the same distribution as the training data (blue bars), and corrupted, out-of-distribution test data (orange bars). The plots show the results using the segmentation metrics DSC (top), ASD (middle), and HD (bottom). On the within-distribution test data all metrics indicate good segmentation accuracy for all approaches and across different regularization weights. For the out-of-distribution data, Atlas-ISTNs with test time refinement out-perform the ITN and the 1-pass prediction by a significant margin with good robustness to the selection of the regularization weight $\lambda$.}
    \label{fig:synth2d-results}
\end{figure}

\paragraph{Quantitative Results}
We also present quantitative results over the 100 cases for both the clean and the corrupted test sets, summarized in Fig.~\ref{fig:synth2d-results}. Metrics used to evaluate the segmentation performance include Dice similarity coefficient (DSC), average surface distance (ASD) in pixels and Hausdorff distance (HD) in pixels. We observe that for the clean data (blue bars) highly accurate segmentations are obtained with all approaches and across the range of values for the regularization weight $\lambda$, indicated by high DSC and low surface distances. For the case of corrupted test data (orange bars), we can see the clear benefit of Atlas-ISTNs with test time refinement. We also observe that the results are not very sensitive to the regularization weight.

\subsection{3D Cardiac CCTA}\label{subsec:cardiac_experiments}

\begin{figure*}[!h]
    \centering
    \includegraphics[width=\linewidth]{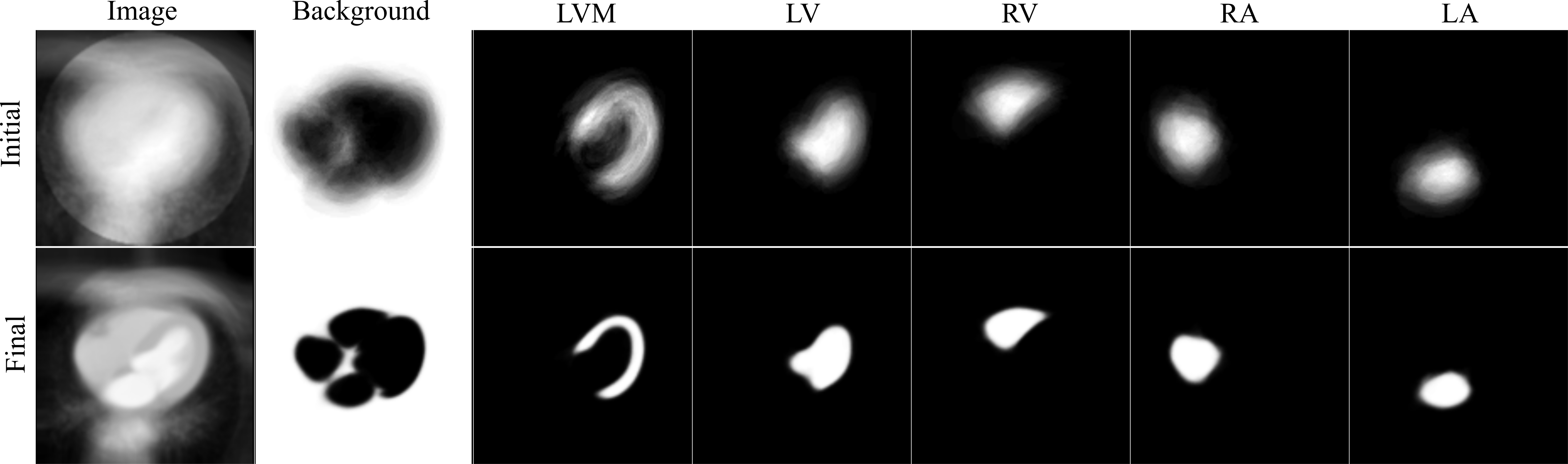}
    \caption{An axial slice through the initial (top row) and final (bottom row) atlas image (first column) and the 6 channels of the atlas labelmap produced when training Atlas-ISTN on multi-label CCTA data.}
    \label{fig:ccta_atlas_2d}
\end{figure*}

The above experiments illustrate the behaviour and benefits of the Atlas-ISTN framework on 2D synthetic data. In the following, we are testing the framework on 3D medical imaging data. We focus on segmentation of large structures of the heart from 3D CCTA which is an important step in both calculation of derived clinical indices such as strain \citep{Nicol2019}, as well as providing a computational domain and boundary conditions for simulations of cardiac function and coronary flow \citep{Taylor2013,Chabiniok2016}. Analysis of Atlas-ISTN is performed with multi-channel labelmaps using a real-world dataset, where ablation studies are performed removing or modifying model components and loss terms from the framework, allowing comparison to baseline segmentation and registration models.

\paragraph{Data Description} 1,109 3D CCTA images from multiple sites around the world were used, consisting of commercial cases which were received by HeartFlow, Inc. for $\text{FFR}_{\text{CT}}$ analysis \citep{Taylor2013}. All cases were processed through the $\text{FFR}_{\text{CT}}$ production pipeline, one output of which is the segmentation of the left ventricle myocardium (LVM). Segmentation of the LVM involves the manual inspection and correction of an automated segmentation produced by a Random Forest + shape fitting model. 109 high quality images were selected for further 3D annotation of the large structures of the heart using in-house tools. In addition to the LVM, these included the left ventricle blood pool (LV), the right ventricle blood pool (RV), the right atrial blood pool (RA) and the left atrial blood pool (LA).

Image intensities were clipped to [-1000, 1000] Hounsfield units, and linearly rescaled to the range [-0.5, 0.5]. All images were isotropically resampled to ensure through-plane resolution (in the $z$ direction) matched the already isotropic in-plane resolution, and images were subsequently downsampled by a factor of 4. While in-plane dimensions are fixed, images were either padded (for the vast majority of cases) or cropped in the $z$-dimension\footnote{SoI were used to define the crop region for these cases.} to obtain volumes of size $128 \times 128 \times 128$. 

The 109 cases with large structure annotations (LSA) were randomly split into 80 training, 10 validation, and 19 testing cases. The remaining 1000 cases with just the LVM were used for additional testing to provide a more diverse test set on which to assess the performance of Atlas-ISTN. In all experiments, Atlas-ISTN is trained using the same 80 cases with all labelled structures, unless otherwise stated.

\paragraph{Model Settings}

PyTorch \citep{NEURIPS2019_9015} was used for all model development. Models were trained for 800 epochs\footnote{Until convergence on validation data.} (or 1200 epochs when using on-the-fly data augmentation) using NVIDIA Tesla V100 SXM2 GPUs (with 32GB memory). The full Atlas-ISTN model was trained with mini-batch size of 8, requiring $\approx$22GB of GPU memory. The Adam optimizer was used with a learning rate of \num{1e-3}, and an exponential learning rate decay with a half-life of 500 epochs was used as this improved model performance. Weighting variables for the training loss in Eq. \eqref{eq:training_loss} were set to $\omega = 1$ and $\lambda = 800$. The atlas update rate was $\eta=0.01$ in Eq. \eqref{eq:atlas_update}. It was found that introducing the affine component later in training stabilized the initial phase of atlas updates, so the affine component was introduced after 200 epochs. Parameter values for the refinement loss in Eq. \eqref{eq:refine_loss} were $\beta^* = 1$, $\gamma^* = 0$ and $\lambda^* = 800$, and number of refinement iterations $K=100$. Refinement often would reach convergence within 30-50 iterations, but 100 iterations were used to ensure all cases had reached convergence. Refinement, performed with a single image at a time, required $\approx$3GB of GPU memory and $\approx$20s runtime for 100 iterations.

\paragraph{Statistical Analyses}

Metrics used to evaluate model performance include Dice similarity coefficient (DSC), average surface distance (ASD) in millimetres and Hausdorff distance (HD) in millimetres. Superiority is shown with a one-sided paired hypothesis test at a significance level of $1\%$, where the test statistic is the mean difference of the metric of interest. We checked the rejection of the null hypothesis using the $98\%$ confidence interval of the test statistic which was estimated with percentage bootstrap using 10,000 repetitions.

\begin{figure*}[!hbtp]
    \centering
    \includegraphics[width=\linewidth]{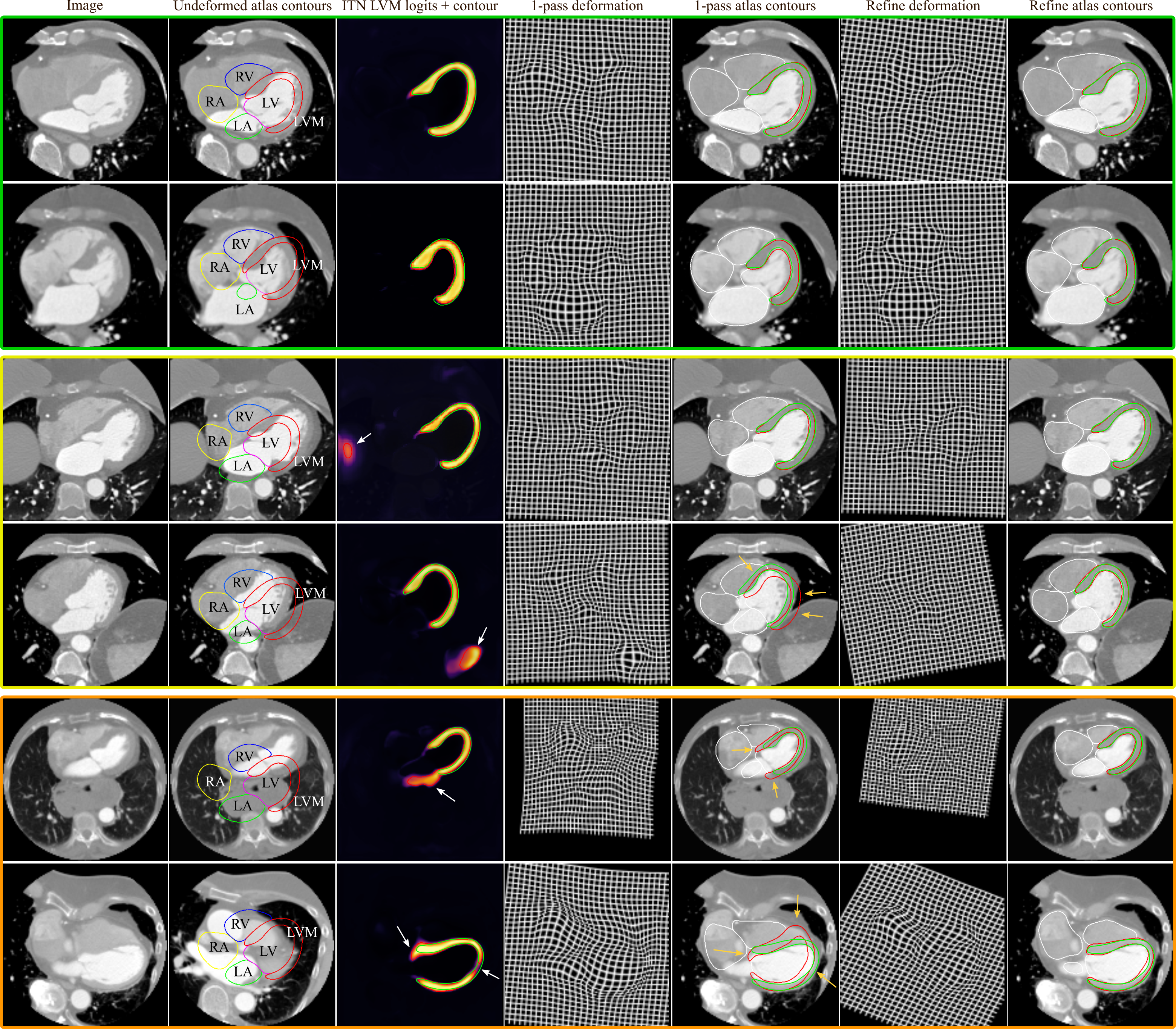}
    \caption{Test cases in order of increasing difficulty (top to bottom), displaying an axial slice from a CCTA image. In columns 3, 5 and 7, manual and predicted contours of the LVM are shown in green and red, respectively, while white contours are used for other predicted structures for clarity of the LVM comparison. Column descriptions are provided above. White arrows highlight false positive and false negative LVM predictions from the ITN. Orange arrows highlight errors between the 1-pass deformed LVM and the manual LVM contours. Rows 1-2 contain examples for which ITN, 1-pass and refinement predictions all perform well, representative of a significant proportion of test cases. Rows 3-4 show cases where the ITN produces spurious segmentations, but refinement is able to circumvent this. In row 4 however, the 1-pass LVM contour is visibly offset, while the refinement produces a better fit. Rows 5-6 show challenging cases, where row 5 contains an example where the heart is particularly small in the field-of-view. On top of this, the ITN predicts a large spurious segmentation extending from the base of the LVM. The deformed contours after 1-pass do not fit the target structures accurately, but after refinement the LVM and other chambers are well positioned. Row 6 shows a case which required a significant rotation and translation of the atlas, so much so that in column 2 a different axial slice is shown to include the undeformed atlas. Additionally the LVM wall is particularly thin, where the ITN predicts a hole near the apex, and an over-segmention in the basal septal region. The significant transformation proves challenging for a single pass of the model, resulting in poor alignment of the atlas to the target SoI. The deformed contours after refinement however align well with the target SoI and bridges the ITN's predicted hole in the LVM (albeit not perfectly fitting to the manual LVM contour). Notice for cases which require larger global transformations (rows 3-6), the deformed grid after refinement contains a more noticeable affine transformation, and the non-rigid component appears to deform the grid less compared to the 1-pass.}
    \label{fig:ccta_examples}
\end{figure*}

\paragraph{Comparison with Baseline Segmentation Model}

A U-net baseline model was trained by optimizing only the weights of the ITN, $\mathrm{\mathbf{S}}_{\theta_s}$, with the segmentation loss $L_s$. The Atlas-ISTN was trained by optimizing jointly the weights of the ITN, $\mathrm{\mathbf{S}}_{\theta_s}$, as well as the STN, $\mathrm{\mathbf{D}}_{\theta_d}$, with all losses in Eq. \eqref{eq:training_loss}. A comparison is made between the U-net and (1) the prediction of the ITN (with identical architecture as the U-net) trained within the Atlas-ISTN framework (`ITN'), (2) the warped atlas labelmap predicted from the first pass of the Atlas-ISTN (`1-pass'), and (3) the warped atlas labelmap after test time refinement of the STN weights (`Refine'). 

Figure \ref{fig:ccta_atlas_2d} shows the initial and final atlas image and multi-channel labelmap resulting from training Atlas-ISTN. The SoI in the final atlas image and labelmap are noticeably sharper, while the background structures in the atlas image remain fairly homogeneous. This is to be expected given that we are optimizing for the segmentation and alignment of the structures depicted in the labelmap. The final atlas image and labelmap are also shown in 3D in Fig. \ref{fig:volumes}.

\begin{table*}
\centering
\begin{tabular*}{0.8\textwidth}{c@{\extracolsep{\fill}}l|ccccc} 
\toprule
\multicolumn{2}{c}{} & \multicolumn{5}{c}{No augmentation}  \\
\cmidrule(lr){3-7}
\multicolumn{4}{c}{} & \multicolumn{3}{c}{Atlas-ISTN}  \\
\cmidrule(lr){5-7}
{Label } & {Metric} & {U-net} & {U-net$_{1cc}$} & {ITN} & {1-pass} & {Refine} \\
\midrule
 & {$\uparrow$ DSC} & 0.883$^{*}$ & 0.883$^{*}$ & \textbf{0.893} & 0.803$^{* \dagger}$ & \underline{\textbf{0.894}} \\
{$LVM$} & {$\downarrow$ ASD} & 0.202$^{* \dagger}$ & 0.190$^{*}$ & \textbf{0.169} & 0.401$^{* \dagger}$ & \underline{\textbf{0.165}} \\
& {$\downarrow$ HD} & 16.401$^{* \dagger}$ & \textbf{6.260}$^{*}$ & 6.842$^{*}$ & 7.775$^{* \dagger}$ & \underline{\textbf{5.255}} \\
\midrule
& {$\uparrow$ DSC} & 0.936$^{*}$ & 0.936$^{*}$ & \textbf{0.941} & 0.896$^{* \dagger}$ & \underline{\textbf{0.943}} \\
{$LV$} & {$\downarrow$ ASD} & 0.123 & 0.123 & \textbf{0.113} & 0.235$^{* \dagger}$ & \underline{\textbf{0.109}} \\
& {$\downarrow$ HD} & 7.157 & 7.132 & \textbf{7.124} & 8.619$^{* \dagger}$ & \underline{\textbf{6.938}} \\
\midrule
& {$\uparrow$ DSC} & 0.894 & 0.895 & \underline{\textbf{0.900}} & 0.846$^{* \dagger}$ & \textbf{0.898}\\
{$RV$} & {$\downarrow$ ASD} & 0.344$^{* \dagger}$ & \textbf{0.287} & 0.309 & 0.483$^{* \dagger}$ & \underline{\textbf{0.284}}\\
& {$\downarrow$ HD} & 29.082$^{* \dagger}$ & \textbf{10.773} & 38.093$^{* \dagger}$ & 12.224$^{* \dagger}$ & \underline{\textbf{10.683}}\\
\midrule
& {$\uparrow$ DSC} & \underline{\textbf{0.862}} & \underline{\textbf{0.862}} & 0.857 & 0.825$^{* \dagger}$ & 0.860\\
{$RA$} & {$\downarrow$ ASD} & \textbf{0.363} & \underline{\textbf{0.362}} & 0.511$^{* \dagger}$ & 0.521$^{* \dagger}$ & 0.383\\
& {$\downarrow$ HD} & 17.263 & \textbf{13.459} & 35.670$^{* \dagger}$ & 13.736 & \underline{\textbf{13.082}}\\
\midrule
& {$\uparrow$ DSC} & 0.886$^{* \dagger}$ & 0.886$^{* \dagger}$ & \textbf{0.899} & 0.846$^{* \dagger}$ & \underline{\textbf{0.900}}\\
{$LA$} & {$\downarrow$ ASD} & 0.344$^{*}$ & 0.338$^{*}$ & \textbf{0.304} & 0.494$^{* \dagger}$ & \underline{\textbf{0.286}} \\
& {$\downarrow$ HD} & 15.149$^{* \dagger}$ & \textbf{12.647}$^{*}$ & 23.031$^{* \dagger}$ & 13.396$^{*}$ & \underline{\textbf{11.725}}\\
\midrule
\multicolumn{2}{c}{} & \multicolumn{5}{c}{With augmentation}  \\
\cmidrule(lr){3-7}
\multicolumn{4}{c}{} & \multicolumn{3}{c}{Atlas-ISTN}  \\
\cmidrule(lr){5-7}
{Label } & {Metric} & {U-net} & {U-net$_{1cc}$} & {ITN} & {1-pass} & {Refine} \\
\midrule
& {$\uparrow$ DSC} & \textbf{0.911} & \underline{\textbf{0.911}} & 0.909 & 0.896$^{* \dagger}$ & 0.911\\
{$LVM$} & {$\downarrow$ ASD} & 0.137 & \textbf{0.136} & 0.138 & 0.169$^{* \dagger}$ & \underline{\textbf{0.136}}\\
& {$\downarrow$ HD} & 6.150 & 4.862 & \textbf{4.785} & 5.313$^{* \dagger}$ & \underline{\textbf{4.544}}\\
\midrule
& {$\uparrow$ DSC} & \textbf{0.950} & \textbf{0.950} & 0.948$^{*}$ & 0.942$^{* \dagger}$ & \underline{\textbf{0.950}}\\
{$LV$} & {$\downarrow$ ASD} & 0.091 & 0.091 & \textbf{0.091} & 0.108$^{* \dagger}$ & \underline{\textbf{0.089}}\\
& {$\downarrow$ HD} & 6.283 & 6.283 & \textbf{5.981} & 6.527$^{*}$ & \underline{\textbf{5.973}}\\
\midrule
& {$\uparrow$ DSC} & 0.903 & 0.903 & \underline{\textbf{0.906}} & 0.897$^{* \dagger}$ & \textbf{0.906}\\
{$RV$} & {$\downarrow$ ASD} & 0.267 & 0.267 &\textbf{0.263} & 0.270 &  \underline{\textbf{0.258}}\\
& {$\downarrow$ HD} & 13.034 & 10.879 & 11.793$^{* \dagger}$ & \underline{\textbf{10.269}} & \textbf{10.647}\\
\midrule
& {$\uparrow$ DSC} & 0.883 & 0.883 & \underline{\textbf{0.884}} & 0.873$^{* \dagger}$ & \textbf{0.883}\\
{$RA$} & {$\downarrow$ ASD} & 0.292 & 0.291 & \underline{\textbf{0.288}} & 0.313$^{* \dagger}$ & \underline{\textbf{0.288}}\\
& {$\downarrow$ HD} & 14.593$^{* \dagger}$ & \textbf{12.243} & 12.862$^{* \dagger}$ & 12.468 & \underline{\textbf{12.187}}\\
\midrule
& {$\uparrow$ DSC} & 0.911 & 0.911 & \underline{\textbf{0.917}} & 0.892$^{* \dagger}$ & \textbf{0.913}\\
{$LA$} & {$\downarrow$ ASD} & \textbf{0.236} & \textbf{0.236} & \underline{\textbf{0.230}} & 0.297$^{* \dagger}$ & 0.238\\
& {$\downarrow$ HD} & \textbf{11.182} & \textbf{11.182} & 12.032 & 11.740$^{*}$ & \underline{\textbf{11.037}}\\
\bottomrule  
\end{tabular*}  
\caption{Comparison with U-net baseline with and without spatial augmentation on the high quality 19 case LSA test set. Arrows indicate direction of metric improvement. Bold numbers are the best and second best, with the best also underlined, for a given metric and augmentation setting. Note that U-net and U-net$_{1cc}$ models with augmentation are identical for LVM and LA labels. Statistically significant ($p < 0.01$) improvement of the best or second best model over a given model is indicated by superscripts $*$ and $\dagger$, respectively.}
\label{tab:baseline_segmentation_19cases}
\end{table*}

Models were initially trained with no data augmentation, and a significant drop in performance was observed from the ITN to 1-pass DSC (Table \ref{tab:baseline_segmentation_19cases}, left). This highlighted the importance of incorporating spatial augmentations when training the Atlas-ISTN with a limited dataset. The ITN learns both global and local image features, and can primarily rely on local image features to make accurate voxel-wise predictions. The STN performance however depends more heavily on learning global image features, given that the predicted displacement field must operate across the entire image to transform SoI to significantly different orientations, scales and morphological configurations. The bottom two rows of Figure \ref{fig:ccta_examples} show the outputs of Atlas-ISTN on particularly challenging cases, where significant global and local deformations are required to register the undeformed atlas to the target SoI. The ITN and 1-pass results are inadequate, while the refinement suitably fits the atlas labelmap to the target SoI. 

Given the limited set of training data, the U-net and the Atlas-ISTN models were also trained with on-the-fly spatial augmentations, which included translation (range: -8 to +8 voxels), rotation (range: -15 to 15 degrees in $x$, $y$, $z$) and scaling (range: 0.9 to 1.1 image \textbf{{\textbf{}}}resolution). 

False positives, or spurious segmentations, are commonly observed in the predictions of voxel-wise segmentation models \citep{Kamnitsas2017,Oktay2018,Larrazabal2019}. A simple and commonly used post-processing step of retaining only the largest connected component of the U-net prediction (`U-net$_{1cc}$') was used as an additional comparison.

\begin{figure}[!h]
    \centering
    \includegraphics[width=\linewidth]{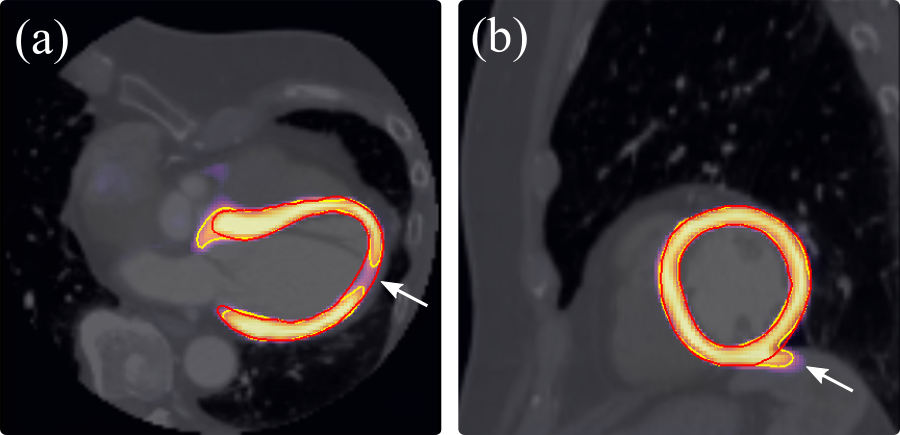}
    \caption{Axial (a) and sagittal (b) planes showing examples where refinement corrects for (a) holes (false negatives) and (b) spurious segmentations (false positives) in the LVM channel predicted by the ITN. Red contour: LVM label of deformed atlas labelmap after refinement. Yellow contour: ITN prediction of the LVM channel. Heatmap: ITN logits of LVM channel. In both (a) and (b), the ITN prediction contour consists of only a single connected component of the LVM.}
    \label{fig:holes_spurious}
\end{figure}

Table \ref{tab:baseline_segmentation_19cases} shows the results of the U-net and Atlas-ISTN using the high quality 19 case LSA test set, for which labels of all chambers were available. For models trained with no data augmentation, we observe slight improvements of the ITN over the U-net for LVM, LV and RV DSC ($1.1\%$, $0.7\%$ and $0.6\%$, respectively), but not for other structures. The 1-pass performance of Atlas-ISTN falls short of the ITN across all metrics for the aforementioned reasons. Refinement produces the best results across almost all metrics, and although DSC improves by a moderate $0.1-0.8\%$ over the ITN, performance on ASD and HD metrics improves significantly. U-net$_{1cc}$ also improves over the U-net in terms of HD and to a lesser extent ASD, but is almost always out-performed by refinement. Fig. \ref{fig:holes_spurious} shows examples where U-net$_{1cc}$ is unable to correct certain false positive and false negative predictions, while refinement of Atlas-ISTN does.

When models are trained with on-the-fly data augmentation, there is a significant improvement across all metrics compared to models trained without augmentation. Most marked is the improvement in 1-pass performance, where for example the 1-pass DSC improves across all channels in absolute terms by between $4.3-8.7\%$, while the refinement DSC improves by $0.8-1.7\%$, leaving a smaller gap between 1-pass and refine metrics. With data augmentation, the U-net, ITN and refine metrics all improve slightly and become more similar to each other. Given that the images in this test set were selected for their high quality, all models produce accurate results and it is perhaps unsurprising that we see only modest improvement with Atlas-ISTN.

To assess performance on a more diverse dataset originating from a wide range of scanners and sites, the same models are run on 1000 test cases each containing a 3D annotation of just the LVM. The results are summarized in Table \ref{tab:baseline_segmentation_1000cases}. All metrics are noticeably worse compared to the LVM metrics on the 19 high quality test cases in Table \ref{tab:baseline_segmentation_19cases} as a reflection of the more diverse and challenging images in the 1000 case test set. For the models trained without data augmentation, the ITN shows an improvement over the U-net across all metrics, and refinement further improves on all metrics. Data augmentation during training improves all metrics significantly, most noticeably for the result of 1-pass DSC with an increase in absolute terms of $14.8\%$ compared to an increase of $2.8\%$ for the refinement DSC.

\begin{table*}
\centering
\begin{tabular*}{\textwidth}{@{\extracolsep{\fill}}l|ccccc|ccccc}
\toprule
\multicolumn{1}{c}{} & \multicolumn{5}{c}{No augmentation} & \multicolumn{5}{c}{With augmentation} \\
\cmidrule(lr){2-6} \cmidrule(lr){7-11}
\multicolumn{3}{c}{} & \multicolumn{3}{c}{Atlas-ISTN} & & & \multicolumn{3}{c}{Atlas-ISTN} \\
\cmidrule(lr){2-6} \cmidrule(lr){7-11}
{Metric} & {U-net} & {U-net$_{1cc}$} & {ITN} & {1-pass} & {Refine} & {U-net} & {U-net$_{1cc}$} & {ITN} & {1-pass} & {Refine} \\
\midrule  
{$\uparrow$ DSC} & 0.840$^{* \dagger}$ & 0.850$^{* \dagger}$ & \textbf{0.863}$^{*}$ & 0.683$^{* \dagger}$ & \underline{\textbf{0.869}} & 0.884$^{* \dagger}$ & \textbf{0.885}$^{*}$ & 0.883$^{* \dagger}$ & 0.850$^{* \dagger}$ & \underline{\textbf{0.888}}\\
{$\downarrow$ ASD} & 0.973$^{* \dagger}$ & 0.417$^{*}$ & \textbf{0.367}$^{*}$ & 1.207$^{* \dagger}$ & \underline{\textbf{0.256}} & 0.301$^{* \dagger}$ & \textbf{0.224}$^{*}$ & 0.342$^{* \dagger}$ & 0.311$^{* \dagger}$ & \underline{\textbf{0.212}}\\
{$\downarrow$ HD} & 38.046$^{* \dagger}$ & \textbf{10.566}$^{*}$ & 22.948$^{* \dagger}$ & 11.763$^{* \dagger}$ & \underline{\textbf{6.120}} & 9.854$^{* \dagger}$ & \textbf{6.440}$^{*}$ & 13.046$^{* \dagger}$ & 6.579$^{*}$ & \underline{\textbf{5.644}}\\
\bottomrule  
\end{tabular*}  
\caption{Comparison with U-net baseline with and without spatial augmentation on the 1000 case LVM test set. Arrows indicate direction of metric improvement. Bold numbers are the best and second best, with the best also underlined, for a given metric and given augmentation setting. Statistically significant ($p < 0.01$) improvement of the best or second best model over a given model is indicated by superscripts $*$ and $\dagger$, respectively.}
\label{tab:baseline_segmentation_1000cases}
\end{table*}

 We observe that U-net$_{1cc}$ improves over the U-net performance both with and without augmentation, and also out-performs the ITN model trained with spatial augmentations. Refinement still produces the best results across both augmentation settings, despite a lower ITN performance compared to the U-net and U-net$_{1cc}$ with augmentation. Additionally, the result of refinement guarantees topology and encourages smoothness of the target structures while U-net$_{1cc}$ does not (see examples in Fig. \ref{fig:holes_spurious}). Statistically significant improvement across all metrics is achieved with refinement compared to all other models.
 
  \begin{figure*}[!h]
    \centering
    \includegraphics[width=\linewidth]{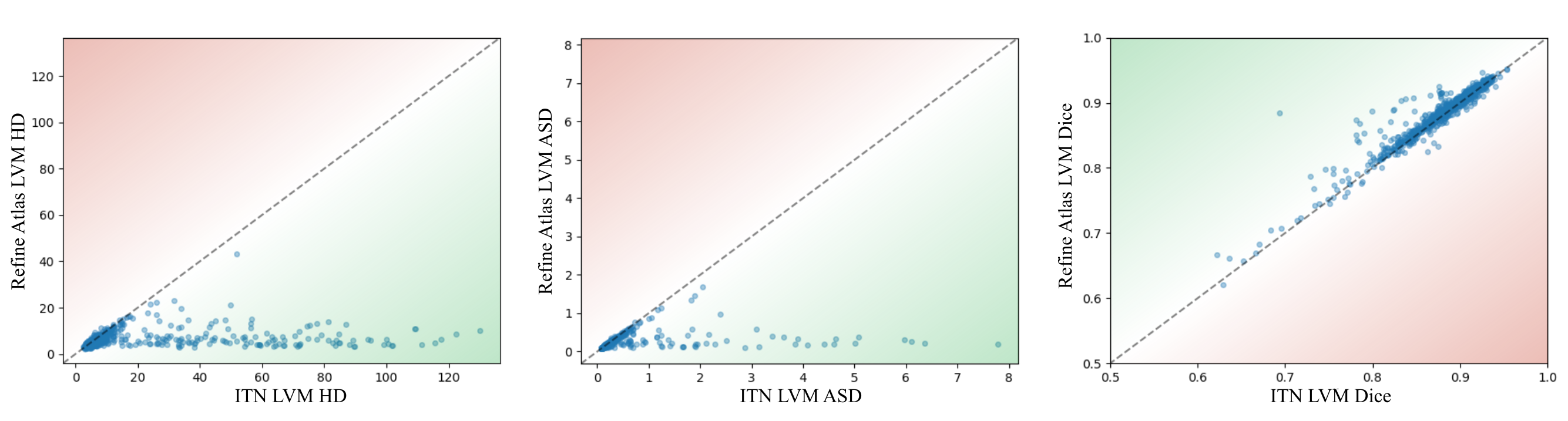}
    \caption{Scatter plots showing HD (left), ASD (middle) and DSC (right) results of Atlas-ISTN on the 1000 case test set, comparing the LVM label from the ITN ($x-$axis) versus refinement ($y$-axis). The green/red gradients indicate increase/decrease in performance with refinement. HD and ASD of the ITN predictions almost always improve with refinement. Degradation observed for some cases in terms of DSC is always small, whereas improvements in DSC can be significant.}
    \label{fig:ccta_scatter_plots}
\end{figure*}

 The improvements achieved with refinement over the ITN prediction are further illustrated in Fig. \ref{fig:ccta_scatter_plots}. Generally we see that there is seldom any degradation in the metrics, while outliers are generally corrected for, particularly for the metrics HD and ASD.

\paragraph{Comparison with Baseline Deformation Models}
 Most related approaches in the literature propose the use of a single pass of an STN at test time to make a registration prediction, which can be used to propagate a segmentation \citep{Balakrishnan2018, Dalca2019, Dong2020}. These methods also use images as inputs to the STN, while Atlas-ISTN enforces an explicit intermediate representation (semantic segmentation) as input to the STN. We make comparisons to variants of the Atlas-ISTN framework, including:

\textbf{VML}: A VoxelMorph-like model - training only an STN (without an ITN), where (i) an intensity image and (ii) the atlas image are passed directly to the STN.

\textbf{Atlas-ISTN$_{-L_s}$}: Atlas-ISTN trained without the segmentation loss $L_s$, thus removing the constraint on the ITN to produce a semantic segmentation as input to the STN.

While Atlas-ISTN$_{-L_s}$ has the same architecture as Atlas-ISTN, VML has slightly fewer than half of the parameters (i.e. just the STN) and takes two single channel images as input. The inputs to the VML model are a target image, $x_i$, (see top left panel in Fig. \ref{fig:volumes} for an example) and the atlas image, $x^a$ (see top right panel in Fig. \ref{fig:volumes} for an example). The atlas labelmap and image are updated as usual (Section \ref{subsec:atlas}). VML is a VoxelMorph-like model similar to \citep{Dalca2019} as it registers a pair of intensity images, where one is an atlas image. In this work the atlas image is constructed on-the-fly during training via registration of training images, while \cite{Dalca2019} explicitly parameterize the atlas image with learnable weights. Also unlike \citep{Dalca2019}, no image-based loss terms are used and ground-truth labelmaps are provided with $100\%$ of the training data for the computation of the loss (referred to as an `auxilliary' loss in \citep{Balakrishnan2018}). Additionally, both an atlas labelmap and image are used during training, where the image is used as an input to the model and the labelmap is used in loss terms Eqs. \eqref{eq:a2s}, \eqref{eq:s2a}. Atlas-ISTN train and test hyper-parameters were kept the same for VML. 

Atlas-ISTN$_{-L_s}$ also uses the same hyper-parameters, although batch-normalization after each convolutional layer in the ITN was required to prevent vanishing gradients during training (given the deeper architecture without $L_s$). Inputs to the STN are the same as for Atlas-ISTN, namely the atlas labelmap and the intermediate representation produced by the ITN. In this case however, the intermediate representation is not enforced to be a semantic segmentation of the SoI. 

An MSE loss between source and target image intensities has been used in previous applications using brain MRI data with VoxelMorph \citep{Balakrishnan2018} and conditional atlases \citep{Dalca2019}. MSE losses between image pairs were experimented with in addition to the loss terms for the labelmaps (Eqs. \eqref{eq:a2s} and \eqref{eq:s2a}), but were found to significantly degrade performance of the models due to the high level of variability in image content, field-of-view, and voxel intensities of structures in the CCTA images, thus were not used in the reported experiments. The final constructed atlas image and labelmap for both of these models were similar in appearance to those produced by Atlas-ISTN.

Table \ref{tab:one_pass_models_1000cases} presents the results of the 1-pass performance of the models trained with data augmentation, with metrics computed for the LVM label using the 1000 case test set. Note, the Atlas-ISTN 1-pass results reported in Table \ref{tab:one_pass_models_1000cases} are repeated from Table \ref{tab:baseline_segmentation_1000cases} for convenience. Atlas-ISTN outperforms both Atlas-ISTN$_{-L_s}$ and VML across all metrics with gaps of about $1.1\%$ and $2.7\%$ in DSC, respectively. Furthermore, Atlas-ISTN$_{-L_s}$ and VML could not benefit from test time refinement of the STN weights by warping the atlas labelmap to the ITN logits as done by Atlas-ISTN, and we found that performing refinement by registering the atlas image to a target intensity image at test time with an MSE loss on image intensities worsened performance with a drop of about $10\%$ in DSC, for the same reasons mentioned earlier. We save the exploration of other image intensity-based losses for future work. 

\begin{table}
\centering
\begin{tabular}{l|ccc}
\toprule
Metric & {VML} & {Atlas-ISTN$_{-L_s}$} & {Atlas-ISTN} \\
 & {1-pass} & {1-pass} & {1-pass} \\
\midrule  
{$\uparrow$ DSC} & 0.822$^{* \dagger}$ & \textbf{0.839}$^{*}$ &\underline{\textbf{0.850}}\\
{$\downarrow$ ASD} & 0.413$^{* \dagger}$ & \textbf{0.368}$^{*}$ & \underline{\textbf{0.311}}\\
{$\downarrow$ HD} & 7.471$^{* \dagger}$ & \textbf{7.302}$^{*}$ & \underline{\textbf{6.579}}\\
\bottomrule  
\end{tabular}  
\caption{Comparison of 1-pass performance of models trained with data augmentation on the 1000 case LVM test set. Bold numbers are the best and second best, with the best also underlined, for a given metric and given augmentation setting. Note further improvement for Atlas-ISTN is achieved with refinement (Table \ref{tab:baseline_segmentation_1000cases}). Statistically significant ($p < 0.01$) improvement of the best or second best model over a given model is indicated by superscripts $*$ and $\dagger$, respectively.}
\label{tab:one_pass_models_1000cases}
\end{table}

\paragraph{Comparison of Framework Variants}

The following variants of Atlas-ISTN were compared:

\textbf{Independent}: Independently trained ITN + refinement of randomly initialized STN weights

\textbf{Fixed}: Atlas-ISTN trained with a fixed atlas

\textbf{SVF}: Atlas-ISTN with SVF-only in the STN 
    
\begin{table*}
\centering
\begin{tabular*}{\textwidth}{@{\extracolsep{\fill}}l|c|ccc|cc|cc|cc}
\toprule
\multicolumn{2}{c}{} & \multicolumn{3}{c}{Independent} & \multicolumn{2}{c}{Fixed} & \multicolumn{2}{c}{SVF} & \multicolumn{2}{c}{Proposed} \\
\cmidrule(lr){3-5} \cmidrule(lr){6-7} \cmidrule(lr){8-9} \cmidrule(lr){10-11}
 {Metric} & {ITN} & {Identity} & {Refine} & {Refine$_{200}$} & {1-pass} & {Refine} & {1-pass} & {Refine} & {1-pass} & {Refine}\\
\midrule  
 {$\uparrow$ DSC} & \textbf{0.884}$^{*}$ & 0.204$^{* \dagger}$ & 0.770$^{* \dagger}$ & 0.820$^{* \dagger}$ & 0.848$^{* \dagger}$ & 0.879$^{* \dagger}$ & 0.854$^{* \dagger}$ & 0.883$^{*}$ & 0.842$^{* \dagger}$ & \underline{\textbf{0.886}}\\
{$\downarrow$ ASD} & 0.301$^{* \dagger}$ & 9.775$^{* \dagger}$ & 1.483$^{* \dagger}$ & 1.035$^{* \dagger}$ & 0.313$^{* \dagger}$ & 0.232$^{* \dagger}$ & 0.298$^{* \dagger}$ & \textbf{0.218}$^{*}$ & 0.328$^{* \dagger}$ & \underline{\textbf{0.213}}\\
{$\downarrow$ HD} & 9.854$^{* \dagger}$ & 32.552$^{* \dagger}$ & 11.312$^{* \dagger}$ & 9.405$^{* \dagger}$ & 6.980$^{* \dagger}$ & 6.196$^{* \dagger}$ & 6.549$^{* \dagger}$ & \textbf{5.539} & 6.567$^{* \dagger}$ & \underline{\textbf{5.506}}\\
\bottomrule  
\end{tabular*}  
\caption{Comparison of 1-pass and refinement results of Atlas-ISTN variants with a U-net baseline on the 1000 case LVM test set, trained with data augmentation. All models use the independently trained U-net as the ITN at test time, where the STN is optimized for each case in refinement. Refine$_{200}$ represents a model where 200 iterations were used in refinement with a randomly initialized STN. Bold numbers are the best and second best, with the best also underlined, for a given metric. Arrows indicate direction of metric improvement. Statistically significant ($p < 0.01$) improvement of the best or second best model over a given model is indicated by superscripts $*$ and $\dagger$, respectively.}
\label{tab:atlas_istn_variants_1000cases}
\end{table*}

Firstly, `Independent' uses an independently trained ITN, followed by test time refinement using an STN with randomly initialized weights and a randomly selected training case for the atlas labelmap\footnote{The randomly selected case is shown on the left in Fig. \ref{fig:volumes}.}. This approach may be useful in a setting where a pre-trained U-net is available, and registration of a training case labelmap to a U-net prediction can be performed by optimizing the STN weights. The `U-net' trained with spatial augmentations (Table \ref{tab:baseline_segmentation_1000cases}) is used for this experiment. The `Fixed' model is Atlas-ISTN trained and tested with an atlas labelmap that is selected from the training data (the same one as for `Independent'), thus by-passing the atlas construction step during training. `SVF' is Atlas-ISTN trained and tested with only the SVF (i.e. no affine component) predicted by the STN. 

These variants allow us to investigate (i) the impact of training the STN before test time refinement, (ii) the impact of learning an unbiased atlas during training, and (iii) the impact of including an affine transformation model in the STN. It was observed that for the U-net and Atlas-ISTN model variants trained with spatial augmentations, ITN performance was generally very similar, and the ITN of a given model could be paired with the STN and atlas labelmap of a different model at test time for refinement without significant differences in 1-pass or refinement performance. In light of this, we substitute the U-net trained independently (with augmentations) for the ITN in all model variants to make a head-to-head comparison of the 1-pass and refinement results, once again using $K=100$ iterations for refinement.

Table \ref{tab:atlas_istn_variants_1000cases} shows the results of the baseline U-net as well as the 1-pass and refinement results of the `Independent', `Fixed', and `SVF' models as well as the Atlas-ISTN. A first pass of the STN for the `Independent' model is meaningless since the STN weights are randomly initialized\footnote{A single pass results in a near identity transform.}, so the column `Identity' shows the metrics between the undeformed fixed atlas labelmap and the test data. Refinement of the STN weights from scratch for `Independent' performs significantly worse than all other models. Standard refinement (with $K=100$) and even refinement with double the iterations ($K=200$) resulted in considerably worse results compared to the other models likely due to falling into bad local minima (e.g. registering to large spurious segmentations near the initial fixed atlas position), or not reaching convergence. The best 1-pass results are obtained with the `SVF' model, followed by `Fixed' and Atlas-ISTN. Refinement with the proposed Atlas-ISTN out-performs all other models, although is closely followed by refinement with `SVF', and both out-perform refinement with `Fixed' across all metrics. This highlights the benefit of using an unbiased atlas labelmap as opposed to using a fixed labelmap. Refinement with Atlas-ISTN, `Fixed', and `SVF' models all improve over the U-net (ITN) in terms of ASD and HD, while Atlas-ISTN also marginally improves DSC. 

We observe that the 1-pass of Atlas-ISTN degrades slightly (by 0.8\% DSC) in this experiment compared to using a jointly trained ITN (Table \ref{tab:baseline_segmentation_1000cases}), although the refinement result overall is quite similar. In practice, the outputs of refinement would be used and not the 1-pass as the final segmentation.

\paragraph{Upper bound LVM model}
To estimate an upper bound on the performance of Atlas-ISTN for the LVM label, an Atlas-ISTN model is trained using 2000 additional training cases which contain only the LVM label. Only the LVM label is used during training, and as a result the constructed atlas does not contain any other foreground labels. The original 80 LSA cases are included in the training process, and the atlas is still constructed from these original 80 cases at the end of each epoch for faster convergence of the atlas construction during training. At each epoch, the 80 LSA cases are passed to the network with on-the-fly spatial augmentation, while another 80 cases are sampled randomly without replacement from the 2000 case dataset without augmentation. The number of epochs was halved so that the models were trained with the same number of iterations as previous models. Hyper-parameters were kept the same as for previously trained models with augmentation, with epoch-dependent parameters adjusted appropriately. U-net and VML models were also trained in this way.

\begin{table}[hb]
\centering
\begin{tabular}{@{} l|ccccc @{}}
\toprule
\multicolumn{1}{c}{} & \multicolumn{2}{c}{} & \multicolumn{3}{c}{Atlas-ISTN} \\
\cmidrule(lr){4-6}
{Metric} & {U-net} & {VML} & {ITN} & {1-pass} & {Refine}\\
\midrule  
{$\uparrow$ DSC}          & 0.911$^{*}$  & 0.890$^{* \dagger}$  & \underline{\textbf{0.918}} & 0.902$^{* \dagger}$ & \textbf{0.911$^{*}$}\\
{$\downarrow$ ASD} & 0.160$^{*}$ & 0.217$^{* \dagger}$ & \underline{\textbf{0.150}} & 0.186$^{* \dagger}$ & \textbf{0.158}$^{*}$\\
{$\downarrow$ HD}          & 5.464$^{*}$ & 6.143$^{* \dagger}$ & \textbf{5.380} & 5.836$^{* \dagger}$ & \underline{\textbf{5.093}}\\
\bottomrule 
\end{tabular}  
\caption{Results on the 1000 case LVM test set of U-net, VML, and Atlas-ISTN models trained with an additional 2000 cases with LVM label only. Bold numbers are the best and second best, with the best also underlined, for a given metric. Statistically significant ($p < 0.01$) improvement of the best or second best model over a given model is indicated by superscripts $*$ and $\dagger$, respectively.}
\label{tab:upper_bound_1000cases}
\end{table}

\begin{figure}[!h]
    \centering
    \includegraphics[width=\linewidth]{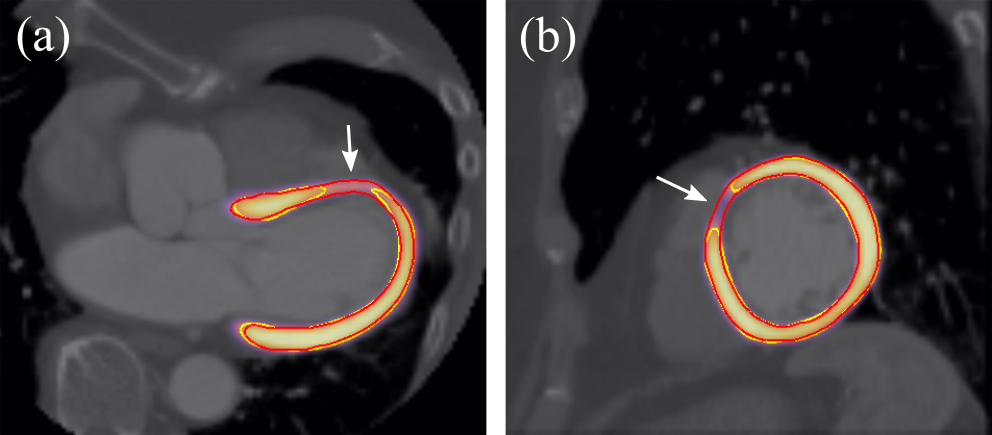}
    \caption{Axial (a) and sagittal (b) planes showing examples where refinement corrects for holes (false negatives) in the LVM channel predicted by the ITN, for the Atlas-ISTN trained with an additional 2000 cases. Red contour: LVM label of deformed atlas labelmap after refinement. Yellow contour: ITN prediction of the LVM channel. Heatmap: ITN logits of LVM channel. In both (a) and (b), the ITN prediction contour consists of only a single connected component of the LVM.}
    \label{fig:holes_lv_model}
\end{figure}

Results on the 1000 case test for these LVM-only models trained with an additional 2000 cases are presented in Table \ref{tab:upper_bound_1000cases}. Interestingly, the ITN out-performs the U-net across all metrics, with a $0.7\%$ DSC increase. The ITN also performs better than the 1-pass and refinement for DSC and ASD. Refinement improves over the ITN, 1-pass and U-net in terms of HD, and performs similarly to the U-net for DSC and ASD. It should be noted that although the ITN out-performs refinement on DSC and ASD, differences between segmentations with DSC $>0.90$ become practically insignificant. Inter-observer DSC for LVM segmentation from cardiac short-axis cine MRI for example is 0.88 \citep{Bai2018}, which one might expect to be slightly higher for segmentation from 3D CCTA images. Also, despite being trained using a significantly larger dataset, the ITN is still prone to predictions with holes, particularly for cases with thin LVM walls, which can be rectified by refinement (Fig. \ref{fig:holes_lv_model}).

Compared to the Atlas-ISTN trained with spatial augmentation on the 80 cases with all structures (Table \ref{tab:baseline_segmentation_1000cases}), an improvement of about $3\%$ in LVM DSC is observed. The improvement in 1-pass performance is about $5\%$ between the two models. The gap in performance between the 1-pass result and refinement is just under $1\%$ DSC for the Atlas-ISTN with the extended training set, compared to a gap of $3.8\%$ for the model trained on 80 cases, demonstrating the effect that a significantly larger training set can have on closing this gap. Finally, while the VML model also significantly improves compared to using limited training samples (Table \ref{tab:one_pass_models_1000cases}), its performance still falls short of the 1-pass of Atlas-ISTN as before.

\paragraph{Inter-Subject Correspondence}

In addition to segmentation, Atlas-ISTN provides correspondence of the SoI across subjects, which can be used for example to propagate the location of anatomical landmarks not originally in the training data, or to assess inter-subject variability. Registrations from subject to atlas and atlas to subject are jointly optimized via a symmetric loss, exploiting the invertibility of the chosen SVF and affine model. The use of this transformation model in conjunction with a mapping to atlas space provides an inherently inverse consistent registration between any two subjects, as described in \citep{joshi2004unbiased}. Specifically, if we consider the registered atlas labelmap (after refinement) as the most accurate estimate of the SoI for each subject at test-time, as demonstrated in the above results, the composition of the transformations to and from atlas space for two subjects provides an inherently inverse consistent mapping of the SoI between these two subjects. Extending Eq. \eqref{eq:refine_transform_atlas}, the mapping of subject $j$ to subject $i$ via atlas space is given by: 

\begin{equation}\label{eq:inter_subject_1}
    y^{a}_i = y^a \circ \Phi^{-1}_i = (y^{a}_j \circ \Phi_j) \circ \Phi^{-1}_i,
\end{equation}

and the mapping of subject $i$ to subject $j$ by:

\begin{equation}\label{eq:inter_subject_2}
    y^{a}_j = y^a \circ \Phi^{-1}_j = (y^{a}_i \circ \Phi_i) \circ \Phi^{-1}_j,
\end{equation}

where $y^a$ is the atlas labelmap, $y^{a}_i$ is the deformed atlas labelmap (representing the SoI) for subject $i$, and $\Phi_i$ is the transformation from subject $i$ to atlas space.

Theoretically, the properties of the diffeomorphic transformation model ensure that these mappings are inverse consistent, though error can arise from numerical precision, the integration of velocity fields, and discrete grid interpolation. Given that the SoI under consideration for a given subject is the deformed atlas labelmap, it follows that if the composed transformation from atlas to subject and back introduces minimal error, then the composition of transformations from one subject to another via atlas space (Eqs. \eqref{eq:inter_subject_1} and \eqref{eq:inter_subject_2}) will similarly have minimal error. 

Two metrics are proposed to estimate the error associated with composing transformations to and from atlas space, including masked inverse consistency error (MICE) and inverse consistency DSC (IC-DSC). An atlas labelmap deformed by both inverse and forward transformations for a given subject $i$ is first defined, $y^a_{IC} = (y^a \circ \Phi^{-1}_i) \circ \Phi_i$. A grid, $G$, of size $N_x \times N_y \times N_z \times 3$, with voxel values corresponding to ($x$, $y$, $z$) voxel indices is also defined along with a twice deformed grid, $G_{IC} = (G \circ \Phi^{-1}_i) \circ \Phi_i$. MICE is the mean absolute displacement error in terms of voxels computed over the voxels masked by the SoI of the undeformed atlas labelmap ($y^a$), i.e. the mean of $\| G - G_{IC} \|$ within the voxels of the atlas SoI\footnote{The SoI voxels are computed by taking the argmax of the atlas labelmap and using the foreground label mask, i.e. voxels $\ge 1$.}. IC-DSC is computed between the atlas labelmap ($y^a$) and the twice deformed atlas labelmap ($y^a_{IC}$). Table \ref{tab:inverse_consistency} shows that MICE is approximately $0.05$ voxels, and IC-DSC is extremely close to 1. This indicates that the predicted SoI resulting from both 1-pass or refinement of Atlas-ISTN for one subject can be mapped to atlas space and subsequently to another subject with minimal error.

\begin{table}[ht!]
\centering
\begin{tabular}{l|cc}
\toprule
\multicolumn{1}{c}{} & \multicolumn{2}{c}{Atlas-ISTN} \\
\cmidrule(lr){2-3}
{Metric} & {1-pass} & {Refine}\\
\midrule  
 {$\uparrow$ IC-DSC} & 0.997 & 0.996\\
 {$\downarrow$ MICE} & 0.0440 & 0.0574\\
\bottomrule 
\end{tabular}  
\caption{Invertibility results on the 19 case test set. IC-DSC: inverse consistency Dice similarity coefficient, averaged over all labels, MICE: masked inverse consistency error (in terms of voxels).}
\label{tab:inverse_consistency}
\end{table}

\subsection{3D Brain MRI} 

While cardiac CT is our main focus in this work, we also conducted experiments for the task of brain structure segmentation in T1-weighted 3D MRI scans. Brain MRI has been the predominant type of data in related work on learning-based image registration \citep{Balakrishnan2018, Dalca2019, Zhao2019, Hoffmann2020}, including our own work on structure-guided image registration \citep{Lee2019a}. Here, we use brain MRI to focus on the specific aspect of generalization across images from different sites and scanners. We train the Atlas-ISTN on data from one site, and compare the segmentation results with and without test time refinement when testing on data from several other sites.

\paragraph{Data Description}

We utilize brain MRI data from three publicly available imaging studies. We use data from the UK Biobank imaging study (UKBB)\footnote{UK Biobank Resource under Application Number 12579} \citep{sudlow2015,miller2016,alfaro-almagro2018}, the Cambridge Centre for Ageing and Neuroscience study (Cam-CAN) \citep{shafto2014cambridge,taylor2017cambridge}, and the IXI dataset\footnote{https://brain-development.org/ixi-dataset/}. Both UKBB and Cam-CAN use a similar imaging protocol with Siemens 3T scanners. IXI contains subsets from three different clinical sites, namely Guy’s Hospital (IXI-Guys) using a Philips 1.5T system, Hammersmith Hospital (IXI-HH) using a Philips 3T scanner, and Institute of Psychiatry (IXI-IoP) using a GE 1.5T system. While the UKBB data is provided with pre-processed images and segmentations, we apply the following pipeline to the Cam-CAN and IXI data in order to match these as closely as possible to UKBB: 1) Skull stripping with ROBEX v1.2\footnote{\url{https://www.nitrc.org/projects/robex}} \citep{iglesias2011robust}; 2) Bias field correction with N4ITK\footnote{\url{https://itk.org}} \citep{tustison2010n4itk}; 3) Sub-cortical brain structure segmentation using FSL FIRST \footnote{\url{https://fsl.fmrib.ox.ac.uk/fsl/fslwiki/FIRST}} \citep{patenaude2011bayesian}. A very similar pipeline had been employed for UKBB with the same automatic segmentation for extracting brain structures. We resample all brain scans to an isotropic 2mm voxel size, and normalize the intensities within the brain masks to zero mean unit variance, where voxels outside the mask are set zero.

We merge the 15 individual brain structures from the FIRST algorithm into a single binary label map, similar to \citep{Lee2019a}. In line with related work on learning-based registration for brain images, we pre-align all scans rigidly to MNI space using drop2\footnote{\url{https://github.com/biomedia-mira/drop2}}, and hence the task of the Atlas-ISTN here is to recover the non-rigid deformation between the images and the to-be-learned brain atlas. We use 100 scans from UKBB for training, 20 for validation, and 200 scans each from UKBB and Cam-CAN and all 581 scans from IXI (with Guys n=322, HH n=185, IoP n=74) for testing the segmentation performance of Atlas-ISTN and baselines.

\paragraph{Model Settings}

Here, an SVF-only transformation model is employed as all scans are rigidly pre-aligned to MNI. The Atlas-ISTN model for brain MRI is based on the same ITN and STN architectures as in the cardiac case with the last scale removed due to the smaller size of the input images. The model is trained for 800 epochs and an exponential learning rate decay with a half-life of 400 epochs. As there is less variation between scans, we found fewer epochs are necessary compared to the cardiac data. Weighting variables for the training loss in Eq. \eqref{eq:training_loss} included $\gamma=1$ as for the cardiac experiments, with changes to $\lambda$ and $\omega$. $\lambda = 500$ was set empirically for brain data. While in cardiac experiments a fixed value of $\omega=1$ was used, for the brain experiments a `fade-in' function was used to initially favour the segmentation loss $L_s$, with weighting on the deformation-related loss terms coming into full effect after about 200 epochs. This was found to slightly improve (observed on the UKBB validation data) the performance of the ITN  and in turn refinement results for the brain experiments.

\begin{equation}
    \omega = \frac{1}{1 + e^{-(t - 200) / 25}},
\end{equation}
  
where $t$ is the epoch index. This approach brought the performance of the ITN closer to that of the U-net, as it possibly reduced the effects of competing gradients from the deformation and segmentation loss terms early in training (competing gradients in multi-task models are studied in \citep{yu2020}). The atlas update rate was also the same as for cardiac, $\eta=0.01$. Parameter values for the refinement loss in Eq.\eqref{eq:refine_loss} were $\beta^* = 1$, $\gamma^* = 0$, $\lambda^* = 500$, and the number of refinement iterations $K=50$.

\paragraph{Baselines}

We evaluate the segmentation accuracy again in terms of DSC, ASD and HD and compare with two baselines, a U-net and a VoxelMorph-like approach (VML) described earlier, i.e. an Atlas-ISTN without the ITN, where an intensity image and the atlas image are passed directly to the STN, by-passing an intermediate representation altogether, and at test time predicting the final transformation in a single forward pass.

\paragraph{Experimental Results}

\begin{figure*}[!h]
    \centering
    \includegraphics[width=\linewidth]{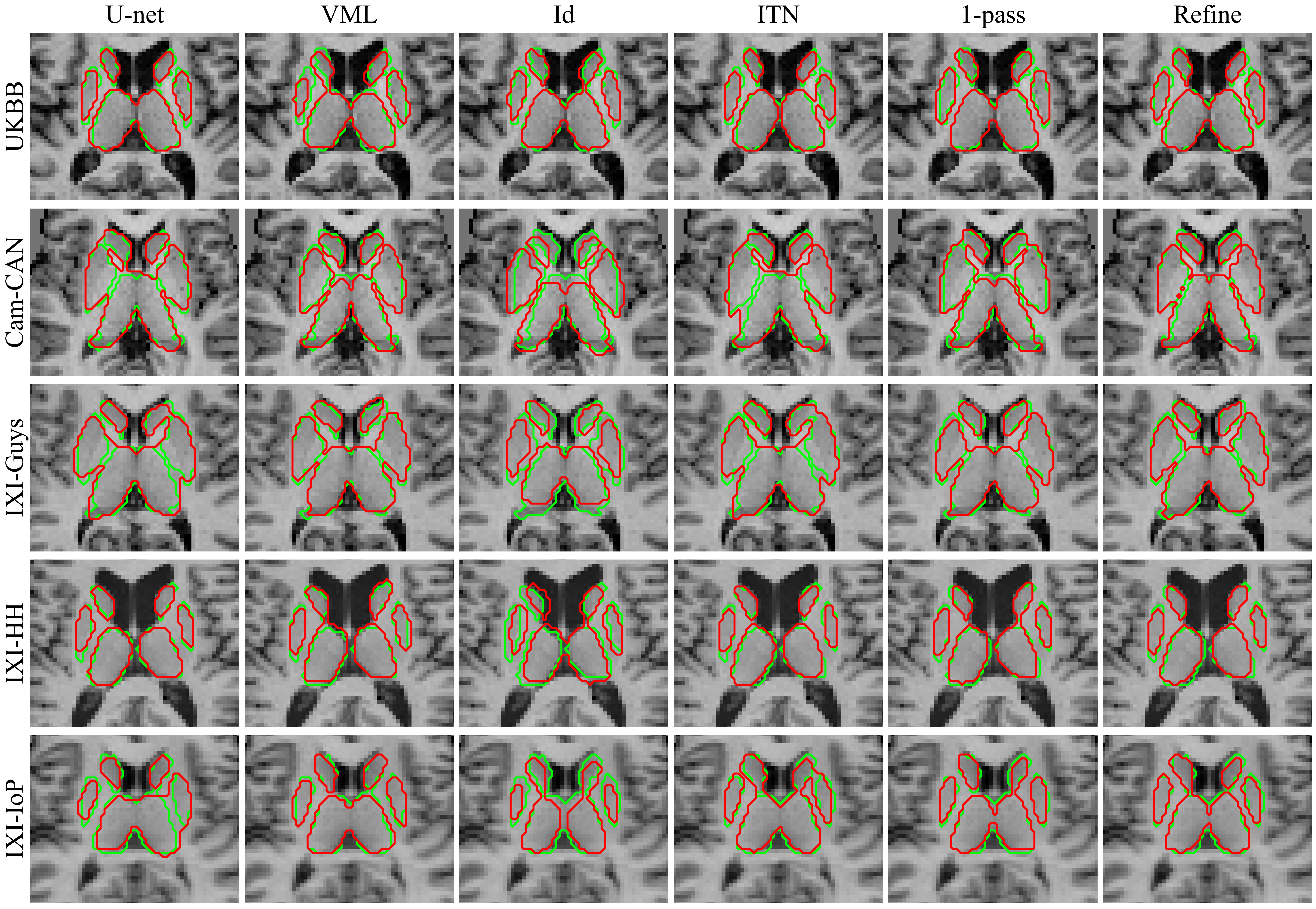}
    \caption{Qualitative results for the brain structure segmentation for fives cases from the five different datasets (one case per row) comparing U-net, VML (i.e. a VoxelMorph-like STN-only baseline), initial atlas alignment `Id', and the three outputs ITN, `1-pass' and `Refine' of our Atlas-ISTN approach. The reference segmentation is displayed with green contours  and the predicted segmentation boundaries with red contours. Note how the voxel-wise segmentation methods, U-net and ITN, tend to merge neighboring structures, while the VML baseline produces less accurate atlas alignment than `1-pass' or `Refine'. The test time refinement of the Atlas-ISTN seems to produce the visually best results, followed by the `1-pass' prediction which acts as an initialization for the refinement.}
    \label{fig:brain_mri}
\end{figure*}

\begin{figure}[!h]
    \centering
    \includegraphics[width=\linewidth]{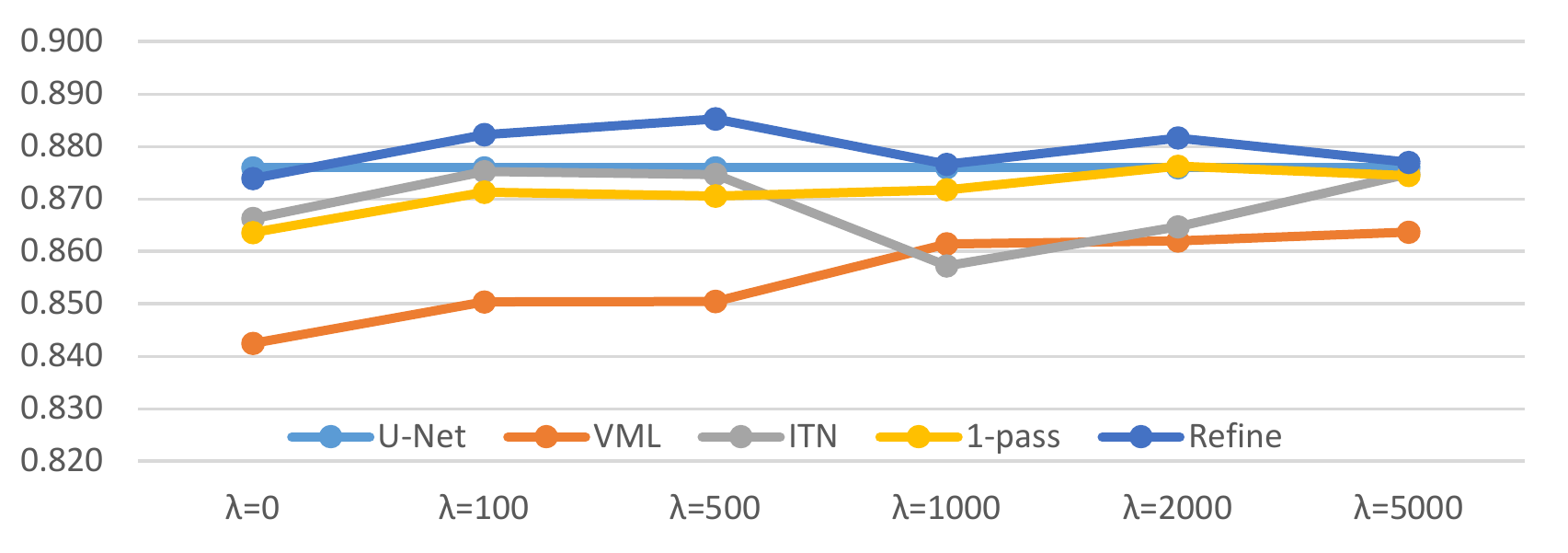}
    \includegraphics[width=\linewidth]{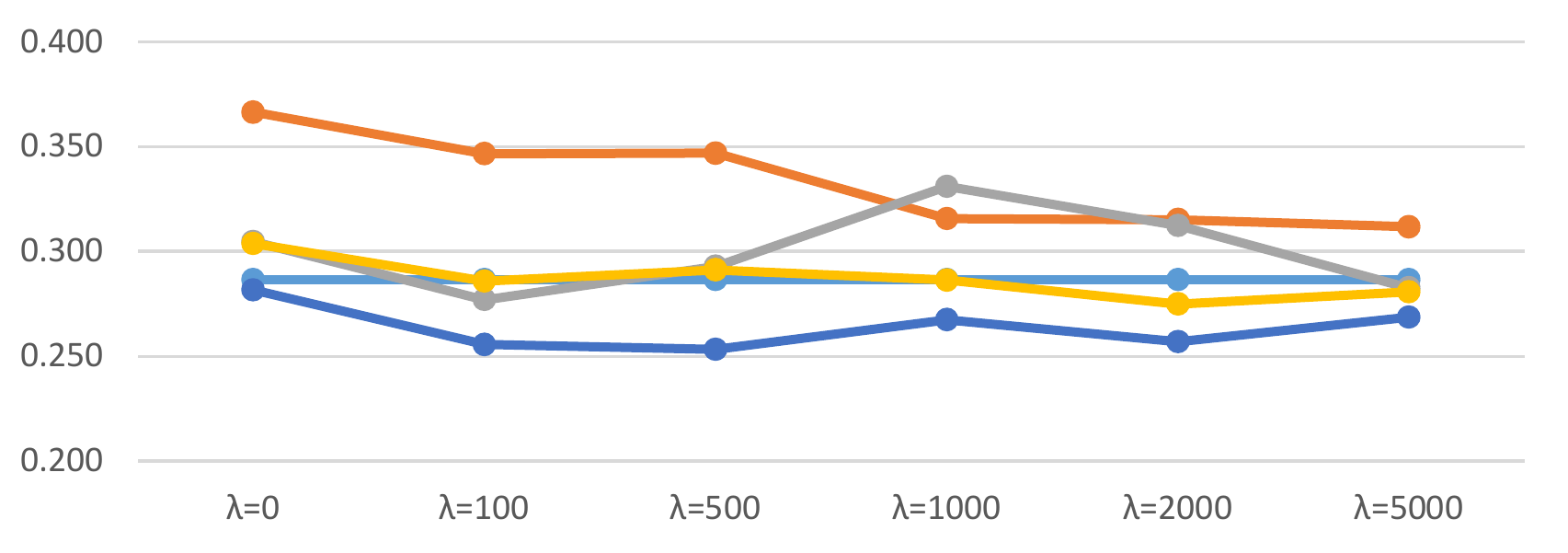}
    \includegraphics[width=\linewidth]{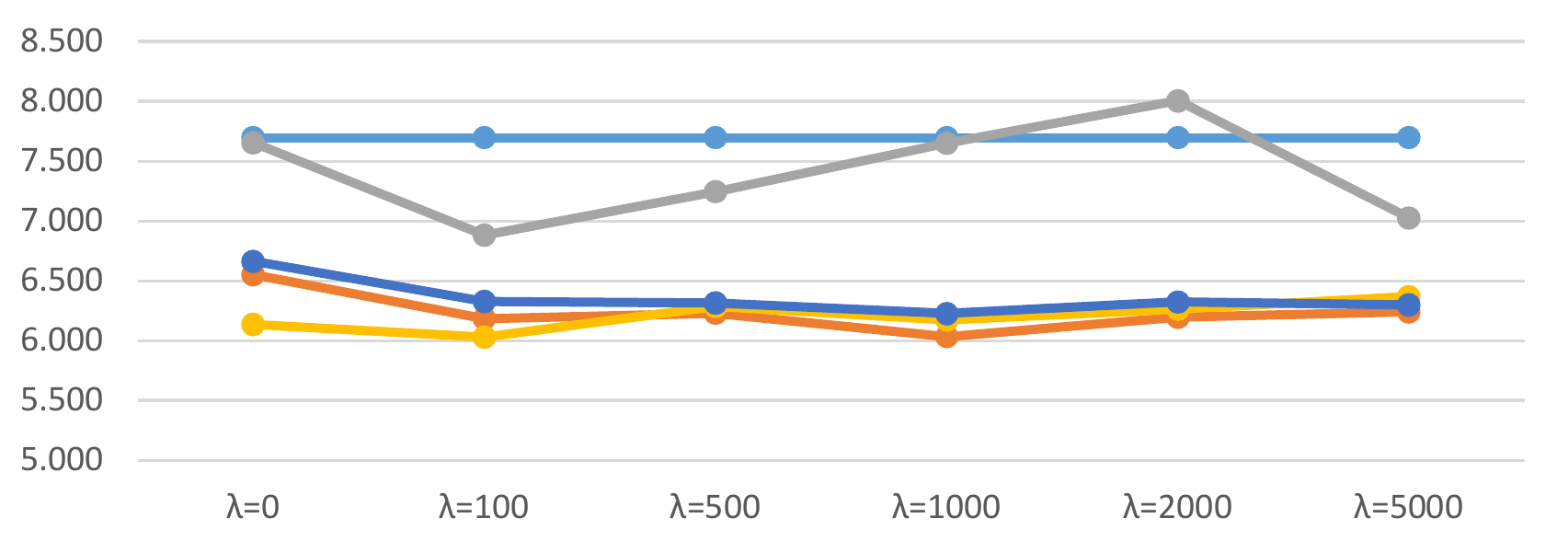}
    \caption{Sensitivity analysis for the effect of the regularization weight $\lambda$ for the application of 3D brain segmentation. The Atlas-ISTN with test time refinement achieves overall best performance for the metrics DSC (top) and ASD (middle) across the range of regularization weights, and performs similar on HD (bottom) compared to the VML baseline and 1-pass prediction. The Atlas-ISTN outperforms the U-net baseline on all three metrics. The results shown here are the averages over the UKBB, Cam-CAN and IXI test datasets. U-net results are independent of $\lambda$ and thus constant.}
    \label{fig:brain_mri_sensitivity}
\end{figure}

The quantitative results are presented in Table~\ref{tab:brain_mri} with a sensitivity analysis regarding the regularization weight $\lambda$ in Fig.~\ref{fig:brain_mri_sensitivity} with visual, qualitative results shown in Fig.~\ref{fig:brain_mri}. Overall, the U-net baseline performed quite well, out-performing the VML baseline on all datasets and performing similarly to the ITN. Atlas-ISTN performed the best on almost all test five datasets across all metrics, with the exception of HD on the Cam-CAN dataset. Test time refinement almost always improved over the ITN and 1-pass performance across all metrics as well. In Fig.~\ref{fig:brain_mri}, the U-net and ITN predictions often include false positives which connect neighbouring structures, most noticeably in rows 2 (Cam-CAN), 3 (IXI-Guys) and 5 (IXI-IoP). These errors are generally corrected with the VML, 1-pass and refinement results.

The VML model under-performed compared to the U-net and Atlas-ISTN models generally on all datasets across all metrics. The VML model did not generalize as well to other datasets, with the performance gap between VML and the other models increasing for datasets less similar to the training dataset. For the least similar dataset compared to the training data, IXI-IoP, the VML model DSC was 0.813 compared to 0.846, 0.839 and 0.862 for the Atlas-ISTN's ITN, 1-pass and refinement results, respectively. The Atlas-ISTN 1-pass results out-performed the VML model across almost all metrics for all datasets, which could be attributed to the intermediate representation provided to the STN. Test time refinement of Atlas-ISTN also produced greater improvements over the ITN and 1-pass results for less similar datasets. On the UKBB dataset, the increase in DSC compared to the ITN and 1-pass were just $0.4\%$ and $0.3\%$, respectively, while for IXI-IoP it was $1.6\%$ and $2.3\%$, respectively.

The sensitivity analysis in Fig.~\ref{fig:brain_mri_sensitivity} shows robustness to the choice of the regularization weight. For DSC and ASD, Atlas-ISTN with test time refinement achieves the best performance for the entire range of $\lambda$ values, while for HD it performs similarly to VML and Atlas-ISTN 1-pass. This also highlights that the improvement for Atlas-ISTN with test time refinement is obtained consistently and independent of the specific strength of regularization.

\begin{table*}
\centering
\begin{tabular*}{0.8\textwidth}{@{\extracolsep{\fill}}c|cccccc} 
\toprule
\multicolumn{4}{c}{} & \multicolumn{3}{c}{Atlas-ISTN} \\
\cmidrule(lr){5-7} 
\multicolumn{1}{c}{Metric} & {U-net} & {VML} & {Id} & {ITN} & {1-pass} & {Refine} \\
\midrule  
\multicolumn{1}{c}{} & \multicolumn{6}{c}{UKBB (n=200)}  \\
\cmidrule(lr){2-7} 
{$\uparrow$ DSC} 	& \textbf{0.900}$^{*}$ & 0.876$^{* \dagger}$ & 0.773$^{* \dagger}$ & 0.898$^{* \dagger}$ & 0.889$^{* \dagger}$ & \underline{\textbf{0.902}}\\
{$\downarrow$ ASD} 	& \textbf{0.208}$^{*}$ & 0.263$^{* \dagger}$ & 0.553$^{* \dagger}$ & 0.213$^{* \dagger}$ & 0.232$^{* \dagger}$ & \underline{\textbf{0.202}}\\
{$\downarrow$ HD} 	& 5.868$^{*}$ & 5.851$^{*}$ & 7.399$^{* \dagger}$ & 5.976$^{* \dagger}$ & \textbf{5.743} & \underline{\textbf{5.651}}\\

\midrule  
\multicolumn{1}{c}{} & \multicolumn{6}{c}{Cam-CAN (n=200)} \\
\cmidrule(lr){2-7} 
{$\uparrow$ DSC}	& 0.869$^{* \dagger}$ & 0.863$^{* \dagger}$ & 0.765$^{* \dagger}$ & 0.866$^{* \dagger}$ & \textbf{0.873} & \underline{\textbf{0.877}}\\
{$\downarrow$ ASD} 	& 0.369$^{* \dagger}$ & 0.315$^{*}$ & 0.597$^{* \dagger}$ & 0.388$^{* \dagger}$ & \underline{\textbf{0.291}} & \textbf{0.301}\\
{$\downarrow$ HD}	& 9.131$^{* \dagger}$ & \underline{\textbf{5.862}} & 7.266$^{* \dagger}$ & 8.715$^{* \dagger}$ & \textbf{6.311}$^{*}$ & 6.926$^{* \dagger}$\\

\midrule
\multicolumn{1}{c}{} & \multicolumn{6}{c}{IXI all (n=581)} \\
\cmidrule(lr){2-7} 
{$\uparrow$ DSC} 	& \textbf{0.880}$^{*}$ & 0.851$^{* \dagger}$ & 0.741$^{* \dagger}$ & 0.880$^{*}$ & 0.874$^{* \dagger}$ & \underline{\textbf{0.890}}\\
{$\downarrow$ ASD} 	& \textbf{0.261}$^{*}$ & 0.343$^{* \dagger}$ & 0.683$^{* \dagger}$ & 0.264$^{*}$ & 0.279$^{* \dagger}$ & \underline{\textbf{0.235}}\\
{$\downarrow$ HD} 	& 7.006$^{* \dagger}$ & 6.104 & 7.610$^{* \dagger}$ & 6.780$^{* \dagger}$ & \textbf{6.075} & \underline{\textbf{6.022}}\\

\midrule
\multicolumn{1}{c}{} & \multicolumn{6}{c}{IXI-Guys (n=322)} \\
\cmidrule(lr){2-7} 
{$\uparrow$ DSC} 	& 0.897$^{*}$ & 0.874$^{* \dagger}$ & 0.769$^{* \dagger}$ & \textbf{0.898}$^{*}$ & 0.890$^{* \dagger}$ & \underline{\textbf{0.906}}\\
{$\downarrow$ ASD}	& 0.217$^{*}$ & 0.278$^{* \dagger}$ & 0.574$^{* \dagger}$ & \textbf{0.215}$^{*}$ & 0.233$^{* \dagger}$ & \underline{\textbf{0.197}}\\
{$\downarrow$ HD} & 5.640$^{*}$ & 5.588$^{*}$ & 6.976$^{* \dagger}$ & 5.779$^{* \dagger}$ & \textbf{5.458}$^{*}$ & \underline{\textbf{5.367}}\\

\midrule
\multicolumn{1}{c}{} & \multicolumn{6}{c}{IXI-HH (n=185)} \\
\cmidrule(lr){2-7} 
{$\uparrow$ DSC} 	& \textbf{0.866}$^{*}$ & 0.827$^{* \dagger}$ & 0.707$^{* \dagger}$ & 0.861$^{* \dagger}$ & 0.859$^{* \dagger}$ & \underline{\textbf{0.875}}\\
{$\downarrow$ ASD} 	& \textbf{0.296}$^{*}$ & 0.403$^{* \dagger}$ & 0.809$^{* \dagger}$ & 0.309$^{* \dagger}$ & 0.310$^{* \dagger}$ & \underline{\textbf{0.269}}\\
{$\downarrow$ HD} 	& 7.945$^{* \dagger}$ & \underline{\textbf{6.418}} & 8.206$^{* \dagger}$ & 7.896$^{* \dagger}$ & \textbf{6.530} & 6.656\\

\midrule
\multicolumn{1}{c}{}  & \multicolumn{6}{c}{IXI-IoP (n=74)}\\
\cmidrule(lr){2-7} 
{$\uparrow$ DSC}	& 0.844$^{*}$ & 0.813$^{* \dagger}$ & 0.708$^{* \dagger}$ & \textbf{0.846}$^{*}$ & 0.839$^{*}$ & \underline{\textbf{0.862}}\\
{$\downarrow$ ASD}	& 0.368$^{*}$ & 0.480$^{* \dagger}$ & 0.845$^{* \dagger}$ & \textbf{0.367}$^{*}$ & 0.401$^{*}$ & \underline{\textbf{0.316}}\\
{$\downarrow$ HD}	& 10.599$^{* \dagger}$ & \textbf{7.568} & 8.882$^{* \dagger}$ & 8.343$^{* \dagger}$ & 7.618 & \underline{\textbf{7.290}}\\

\bottomrule  
\end{tabular*}  
\caption{Quantitative results for the brain MRI experiments where segmentation methods are trained on 100 cases from UKBB and tested on different datasets from three imaging studies, UKBB, Cam-CAN and IXI. The column `Id' refer to the initial alignment (with identity transformation) of the constructed atlas to put the resulting DSC, ASD and HD into context. Bold numbers are the best and second best per row, with the best also being underlined. Statistically significant ($p < 0.01$) improvement of the best or second best model over a given model is indicated by superscripts $*$ and $\dagger$, respectively.}
\label{tab:brain_mri}
\end{table*}

\section{Discussion}

Results from experiments with synthetic 2D data, real 3D CCTA and T1-weighted brain MRI scans illustrate the benefits of the Atlas-ISTN framework.  

Firstly, experiments with synthetic 2D data demonstrated that improvements in terms of DSC, ASD and HD were achieved with test-time refinement on corrupted, out-of-distribution test images compared to both the ITN and 1-pass of Atlas-ISTN, and that this improvement was insensitive to the choice of the weighting ($\lambda$) of the regularization term (Fig.~\ref{fig:synth2d-results}). The parameter $\lambda$ could also be adjusted for test-time refinement to adjust the rigidity of the deformation used to obtain a final registration of the atlas. Since only the non-rigid component is penalized by the regularization, higher values of $\lambda$ result in deformations that rely increasingly on the affine parameters. This flexibility allows for scenarios where one might want to restrict the non-rigid deformation from adhering too closely to potentially noisy predictions of the ITN, and retain more of the underlying shape of the atlas. On clean test data, the Atlas-ISTN performs en-par with a baseline U-net but with the added benefit of yielding atlas correspondences.

For the application to real 3D data, values of $\lambda$ were selected empirically. Generally, values of $\lambda$ that were too high resulted in worse performance with test time refinement, and values that were too low could produce undesirably sharp gradients in the predicted non-rigid deformations and less robustness to spurious segmentations from the ITN. Adjustment of $\lambda$ between training and test time was not extensively explored on the real cardiac data, although the experiments with synthetic data demonstrate the potential benefits.

The experiments with both CCTA and T1-weighted brain MRI data demonstrate the improved performance of Atlas-ISTN over segmentation-only and registration-only baseline models. The 1-pass of Atlas-ISTN out-performed the VML model in both applications, where a larger gap in performance was observed for data further from the training distribution (Table \ref{tab:brain_mri}). The 1-pass of Atlas-ISTN also out-performed the 1-pass of an Atlas-ISTN model without an imposed intermediate representation (Table \ref{tab:one_pass_models_1000cases}). This indicates that the use of semantic segmentations as intermediate representations in Atlas-ISTN are advantageous for the 1-pass registration, providing robustness to variability and noise in the input images. This reinforces the findings of \citep{Lee2019b} which proposed the use of intermediate representations in an ISTN for pairwise registration.

The experiments with synthetic and real data also consistently demonstrate that test time refinement improves performance over a 1-pass of Atlas-ISTN. This improvement was also shown to be larger for data further from the training distribution (Table \ref{tab:brain_mri}). Test time refinement of Atlas-ISTN also consistently out-performed the ITN and baseline U-net in terms of DSC, ASD and HD, and demonstrated the ability to circumvent spurious segmentations and false negatives in the voxel-wise ITN prediction (as shown in Figs. \ref{fig:ccta_examples} and \ref{fig:holes_spurious}). Particularly for data further from the training distribution, test-time refinement produced the greatest improvements over the U-net and ITN, as well as the 1-pass of Atlas-ISTN (Tables \ref{tab:baseline_segmentation_1000cases} and \ref{tab:brain_mri}).

Obtaining and annotating large sets of 3D medical image data is a common challenge. Most of our experiments with real 3D data involved training with datasets of fewer than 100 samples. In experiments where models were trained on a significantly larger dataset of CCTA cases with LVM labels only, the performance gap between 1-pass and test time refinement of Atlas-ISTN narrowed significantly (Table \ref{tab:upper_bound_1000cases}), with ITN and test time refinement of Atlas-ISTN still out-performing the 1-pass result and the VML model. This suggests that learned (1-pass) registration models generally require more training data to reach the performance of the ITN and subsequent test time refinement of Atlas-ISTN, particularly for datasets which may have significant inherent spatial variability like CCTA data. This demonstrates the advantage of using test time refinement with Atlas-ISTN for models trained with a limited size annotated dataset. Furthermore, when auxiliary information in the form of labelmaps is available, it can be used not only in the loss but also to generate intermediate representations.

While hyper-parameters were tuned, an exhaustive search of architectures and parameters was not undertaken. Between the 3D cardiac and brain experiments, minimal changes were made to hyper-parameters, with modifications to $\lambda$ and $\omega$ providing some slight performance gains (Eq. \eqref{eq:training_loss}). A U-net was chosen as a baseline and as the ITN component, but any suitable segmentation model (or image-to-image architecture) can be used as the ITN. Improvements in the ITN performance are also likely to result in improvements in performance of test time refinement, as shown throughout the experiments. 

We do not assess the sensitivity of the constructed atlas to hyper-parameters. Additionally, the atlas construction process does not explicitly guarantee a consistent topology for the atlas labelmap, i.e. an atlas which conforms to the desired target topology, with structures that are smooth, contiguous, non-overlapping, and each containing only a single connected component. However, empirically these properties were observed for all explored settings, both for single label structures with 2D synthetic and 3D MRI data, and in the setting of multiple cardiac structures with 3D CCTA data. The general technique of averaging the labelmaps of multiple structures from a set of co-registered images has after all been used in the past for cardiac \citep{Bai2015} and brain \citep{joshi2004unbiased,Cabezas2011} atlases. For more complex structures than those explored in this work, a more sophisticated fusion step during the training process or after training may help ensure topological consistency of the final atlas labelmap.  

While we propose a method to generate an unbiased atlas by incorporating some ideas from classic atlas construction methods \citep{joshi2004unbiased} into a deep learning framework, we do not explore the use of constraints that can be imposed to ensure a mean shape \citep{joshi2004unbiased,Dalca2019,Bone2020}, or preserve high resolution detail in the atlas image \citep{Guimond2000}. Indeed, we do not extensively explore the use of image loss terms which are typically part of atlas construction frameworks, particularly for atlas images. An MSE loss was used in training Atlas-ISTN with the cardiac CCTA image data, which was found to be poorly suited to the modality and reduced segmentation performance (for reasons discussed in Sec. \ref{subsec:cardiac_experiments}). In this work, improving segmentation accuracy over baseline methods was facilitated by the construction of the atlas labelmap. Additionally, the use of the constructed atlas labelmap in test time refinement was demonstrated to improve performance compared to the use of a fixed atlas labelmap (Table \ref{tab:atlas_istn_variants_1000cases}). We have not explored the option of constructing an atlas using traditional methods for subsequent use in a CNN framework as in \citep{Dong2020}, though the proposed framework arguably reduces the need for this additional step. We leave further exploration of atlas construction within the Atlas-ISTN framework to future work.

Further exploration of image similarity loss terms could open opportunities not just for atlas construction but for semi-supervised learning with CCTA data, as done previously for brain MRI \citep{Balakrishnan2018,Dalca2019,Dalca2019a,Xu2019}, chest X-ray \citep{Mansilla2020} and knee MRI \citep{Xu2019}. Cardiac CCTA presents several unique challenges, including large variability in the extent of visible anatomy, variability in shape of the field-of-view, tissue-level intensity variations due to differences in contrast timing, artifacts due to implants and differences between scanners and acquisition protocols. Accounting for these factors would be important when introducing image similarity losses into the network. As demonstrated with Atlas-ISTN, an intermediate representation of SoI segmentations helps to mitigate for such variations, with the proposed method out-performing the baseline U-net and registration models when assessed on 1000 test cases from a wide range of sites around the world.    

Our findings on brain MRI may further be of interest to people working on learning-based image registration. The fact that the accuracy of the 1-pass predictions became significantly worse for data from a different site might suggest that the same may be true for STN-based registration approaches such as \citep{Balakrishnan2018,Dalca2019a}. In particular, this may be an issue when only limited amounts of training data are available, a point we made in our earlier work \cite{Lee2019a}. 

The Atlas-ISTN framework could also be adapted for other potential applications. Firstly, new structures such as landmarks, or representations such as meshes, can be added directly into the atlas after training. Additional labels can also be learned by the ITN without contributing to the atlas construction (e.g. structures which may not have strong one-to-one mappings between cases, such as coronary trees). Inter-subject correspondence via atlas space also provides the opportunity for population shape and motion analysis.

\section{Conclusions}

Atlas-ISTN provides a framework to jointly learn image segmentation and registration, while simultaneously generating a population-derived atlas used in the model training process. Subsequent registration of the atlas labelmap via test time refinement provides a topologically consistent and accurate segmentation of the target structures. We have demonstrated quantitatively and qualitatively the improvement in segmentation performance of the proposed Atlas-ISTN model over baseline segmentation and registration models. Through ablation studies, we have also demonstrated the importance of different design choices, including the use of both affine and non-rigid components of the transformation model, the value of using intermediate representations of SoI for registration, and the advantage of using an unbiased atlas compared to a fixed atlas. Furthermore, Atlas-ISTN shows greater improvement over segmentation and registration baselines on test data further from the training distribution, particularly when trained with limited data. Atlas-ISTN may benefit segmentation applications where a known topology is expected, and where inter-subject correspondences may be of interest.

\section*{Acknowledgements}
This research was funded by HeartFlow, Inc. 

\bibliographystyle{model2-names.bst}\biboptions{authoryear}
\bibliography{refs}

\end{document}